\documentclass{article}

\usepackage[final,nonatbib]{neurips_2022}

\usepackage[utf8]{inputenc} 
\usepackage[T1]{fontenc}    
\usepackage{hyperref}       
\usepackage{url}            
\usepackage{booktabs}       
\usepackage{amsfonts}       
\usepackage{nicefrac}       
\usepackage{microtype}      
\usepackage{xcolor}         
\usepackage{graphicx}
\usepackage{algorithm}
\usepackage{amsmath}
\usepackage{algpseudocode}
\algrenewcommand\algorithmicrequire{\textbf{Input:}}
\algrenewcommand\algorithmicensure{\textbf{Output:}}
\usepackage{multirow}
\usepackage{chemfig}
\usepackage{caption}
\usepackage{subcaption}
\usepackage{CJK}
\usepackage{blindtext}
\usepackage{todonotes}
\usepackage{paralist}
\usepackage{wrapfig}
\usepackage{lipsum}
\usepackage{pdfpages}

\usepackage[sorting=none]{biblatex}
\newcommand{\rebuttal}[1]{\textcolor{black}{#1}}

\newcommand{\method}{\textsc{Desert}}
\newcommand{\methodligand}{\textsc{Desert-Ligand}}
\newcommand{\methodpocket}{\textsc{Desert-Pocket}}
\newcommand{\methodmodel}{\textsc{Shape2Mol}}
\newcommand\blfootnote[1]{%
\hypersetup{hidelinks}
  \begingroup
  \renewcommand\thefootnote{}\footnote{#1}%
  \addtocounter{footnote}{-1}%
  \endgroup
}

\newcommand{\citetx}[1]{\AtNextCite{\defcounter{maxnames}{1}}\citeauthor{#1} \cite{#1}}

\title{Zero-Shot 3D Drug Design by Sketching and Generating}

\author{
Siyu Long$^{*1}$, Yi Zhou$^{2}$, Xinyu Dai$^{1}$, Hao Zhou$^{3}$\\ $^{1}$National Key Laboratory for Novel Software Technology, Nanjing University\\
$^{2}$ByteDance AI Lab\\
$^{3}$Institute for AI Industry Research (AIR), Tsinghua University\\
{\tt longsy@smail.nju.edu.cn, zhouyi.naive@bytedance.com}\\
{\tt daixinyu@nju.edu.cn, zhouhao@air.tsinghua.edu.cn}
}

\begin{document}
\begin{CJK*}{UTF8}{gbsn}

\maketitle

\begin{abstract}
Drug design is a crucial step in the drug discovery cycle.
Recently, various deep learning-based methods design drugs by generating novel molecules from scratch, avoiding traversing large-scale drug libraries.
However, they depend on scarce experimental data or time-consuming docking simulation, leading to overfitting issues with limited training data and slow generation speed.
In this study, we propose the zero-shot drug design method \method~(\textbf{D}rug d\textbf{E}sign by \textbf{S}k\textbf{E}tching and gene\textbf{R}a\textbf{T}ing).
Specifically, \method~splits the design process into two stages: sketching and generating, and bridges them with the molecular shape.
The two-stage fashion enables our method to utilize the large-scale molecular database to reduce the need for experimental data and docking simulation.
Experiments show that \method~achieves a new state-of-the-art at a fast speed.\footnote{Code is available at https://github.com/longlongman/DESERT.}
\blfootnote{$*$ Work was done when Siyu Long was a research intern at Bytedance AI Lab.}
\end{abstract}

\section{Introduction}
\label{sec:intro}

Drug design is a crucial step in the drug discovery cycle, which is the inventive process of finding new drugs based on a biological target~(usually a protein pocket) \cite{pocket,ShitongLuo,mss}.
However, seeking appropriate drugs for a particular target is quite challenging due to the enormous space of drug candidates~(almost $10^{33}$).
Traditional drug design approaches usually employ virtual screening~\cite{docking1,docking2,docking3} and molecular dynamics~\cite{md1,md2} to traverse in a large scaled drug library, which is time-consuming and could not produce novel drug candidates.
Recently, a line of work proposes to realize drug design by generating drug molecules from scratch using deep generative models~\cite{jtvae,ShitongLuo,3dmars}, which is quite promising due to the fast speed and the ability of \textit{de-novo} drug design.


Most of current drug generation model are developed upon 
1D~(SMILES) 
\cite{string1,string2,string3} 
or 2D~(molecular graph) 
\cite{graph1,graph2,graph3,graphmap,bio,graph4} 
molecular structures, which heavily rely on expensive experimental data for supervised training while ignoring the 3D interaction information between the drug and the pocket.
In a word, they attempt to find a molecule that maximizes the score given by a bio-activity predictor trained on experimental data. 
They employ different optimization methods, such as Generative Adversarial Network (GAN) \cite{molgan}, Bayesian Optimization (BO) \cite{bo1,bo2,jtvae}, Reinforcement Learning (RL) \cite{rational,rl1,molgan,rl2,graphaf,moldqn}, Evolutionary and Genetic Algorithms (EA/GA) \cite{graphga,ga1,ga_d,ga_2,ga_3}, and Markov Chain Monte Carlo (MCMC) \cite{mars,sampling1}, in the molecular space for obtaining desired drug molecule under certain constraints.
However, we argue that a purely data-driven approach of drug design is practically limited, since in most cases, the quantity of experimental drug-pocket pairs hardly enables supervised training of \textit{de-novo} drug design.
Generally, most protein pockets lack bio-activity data, and learning on noisy and deficient data may lead to severe overfitting problems.

\begin{figure}
\centering
\begin{minipage}[t]{0.45\textwidth}
\centering
\centering
\includegraphics[scale=0.32]{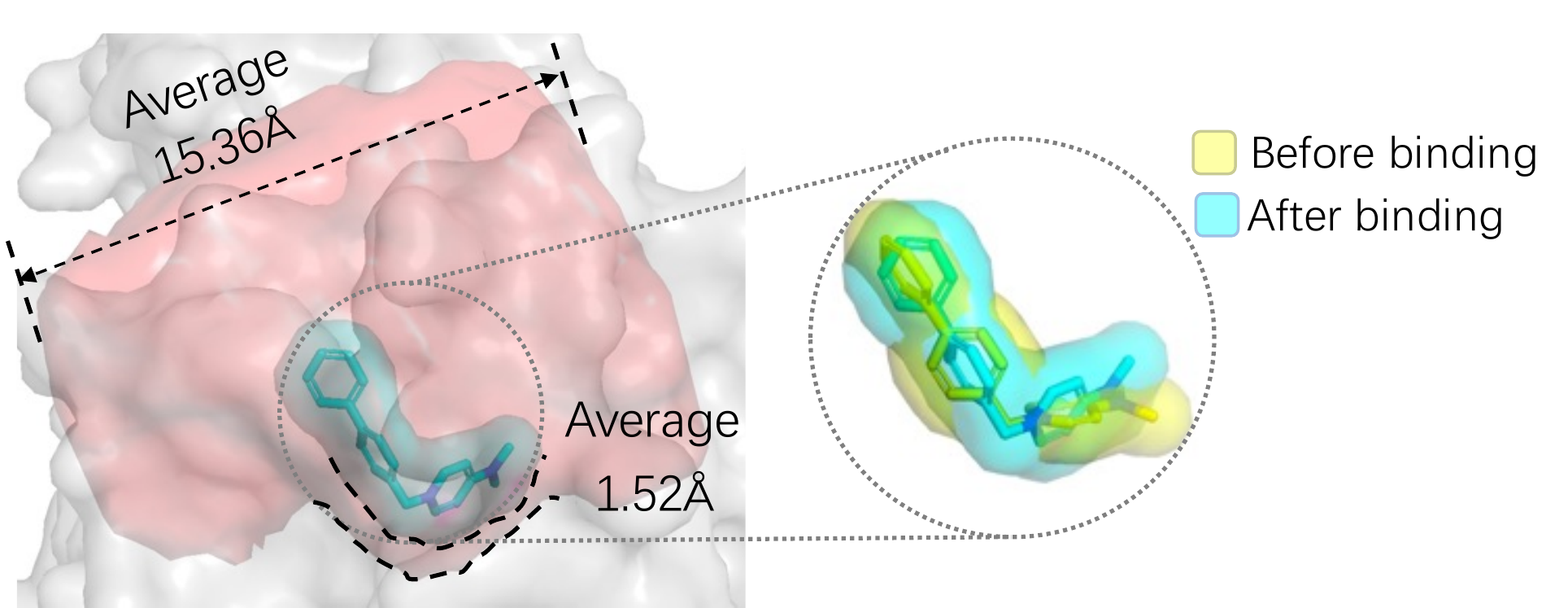}
\caption{When a drug is binding to a pocket, its shape does not change too much (with an RMSD less than $1.391$\AA) and is complementary to the pocket.
}
\label{fig:comchange}
\end{minipage}
\hfill
\begin{minipage}[t]{0.45\textwidth}
\centering
\includegraphics[scale=0.4]{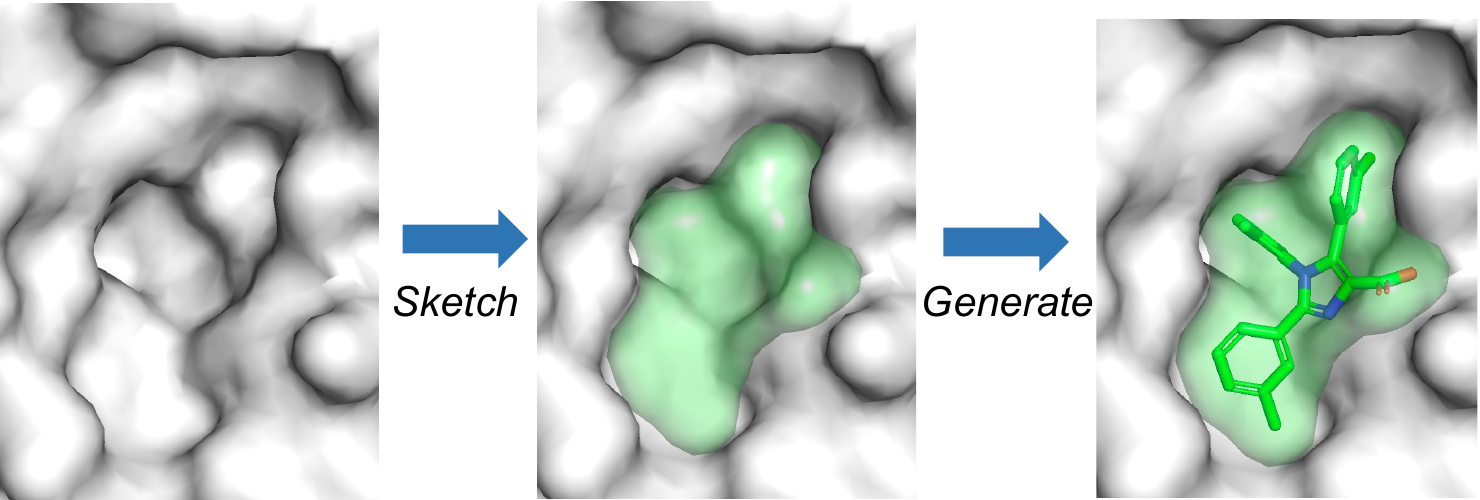}
\caption{\method~splits the whole drug design process into two stages: sketching the shape and generating the molecules.}
\label{fig:mol_design}
\end{minipage}
\end{figure}






Recently, several drug design models have been proposed to directly generate drug molecules in the 3D space, the realistic space of drug-target interaction.
Generating in such space is very promising for the potential of leveraging some prior knowledge~(e.g., physical knowledge) instead of entirely counting on data-driven methodology.
Specifically, \citetx{liGAN} and \citetx{ShitongLuo} propose efficient 3D generative models to learn the atom density conditioned on protein pockets, with GAN and auto-regressive models, respectively. Nevertheless experimental data are still obligatory in their models for achieving satisfactory results.
Even more noteworthy is GEKO~\cite{3dmars}, which combines the ideas of 3D generation and physical simulation to obtain state-of-the-art drug design performance without the help of large-scale experimental data.
Intuitively, GEKO performs geometric editing in the 3D molecule space guided by docking simulation~(physical knowledge) \cite{vina, vina2}. 
However, GEKO may suffer from two concerns: a) the frequent invocation of docking is very time consuming, which significantly slows down the speed of the drug design model~\cite{docktime,docktime1}. b) docking may not always be accurate enough, especially in some complex settings~\cite{dockreliable,dockreliable1}. In such a case, being heavily dependent on the docking accuracy could hurt the generalization of the proposed drug design model.



In this paper, we propose the zero-shot approach \method, namely \textbf{D}rug d\textbf{E}sign by \textbf{S}k\textbf{E}tching and gene\textbf{R}a\textbf{T}ing.
Motivated by the idea of \textit{structure determines properties} \cite{sdp1,sdp2,sdp3,sdp4},
\method~is built on the assumption that molecular shape determines bio-activity between drug molecules and its target pocket.
In other words, we suppose that a drug candidate would have satisfactory bio-activity to a target pocket if their shapes are complementary~(see Figure \ref{fig:comchange}).

With such prior, as shown in Figure \ref{fig:mol_design}, \method~splits the whole drug design process into two stage: sketching and generating, which employs the \textit{molecular shape} as the bridge of the two stages.
Such splitting makes \method~enjoy two advantages:
a) \method~does not heavily rely on docking simulation, which only optionally uses docking for post-process and thus avoids the aforementioned disadvantages of GEKO.
b) \method~abandons the expensive experimental data. Specifically, in the sketching stage, we only need to sample some reasonable shapes complementary to the target pocket. 
In the generating stage, \method~proposes to employ a generative pre-trained model~(from shape to concrete molecular) to fill the shape obtained in the last sketching stage. 
Notably, the generative pre-trained model is only trained on the ZINC database, which contains 1000M pairs of molecules and their corresponding shapes.
This process does not rely on experimental data, making \method~work in a zero-shot fashion. \footnote{The generation stage of \method~can also be equipped with chemistry priors, we conduct some experiments in Appendix 2.2.}

Note that \method~is not baseless. Besides the idea that structure determines properties~(an important concept of  Structural Biochemistry), we also have preliminary results to verify our assumption.
Figure \ref{fig:comchange} shows that after binding to a pocket, the root-mean-square deviation (RMSD) is not too shabby ($1.391$\AA) compared with molecular conformation generation methods such as CGCF \cite{conf1} ($1.248$\AA) and ETKDG \cite{conf2} ($1.042$\AA).
The results suggest that the molecular shape is stable when molecules bind to proteins. We also show in Figure \ref{fig:comchange} that a ligand often attaches tight to a pocket, which means their shape is complementary.

In summary, our contributions are three-fold: (1) We propose \method, a generative  method for \textit{de-novo} 3D shape-based drug design in the zero-shot setting. (2) \method~trains a pre-trained model from massive unbound molecules \footnote{For clarity, the following terms will be used throughout this study: ``bound drug/molecule'' (or ``unbound drug/molecule'') refers to the drug/molecule that is bound (or unbound) to proteins \cite{bound-notation}.}, eliminating the constraints of labeled data. (3) \method~achieves a new state-of-the-art result (the docking score improves $0.79 \mathrm{kcal/mol}$ and $2.93\mathrm{kcal/mol}$ over the best supervised method on two datasets) at a fast speed (about $20$ times faster than GEKO\footnote{We calculate the speed by measuring the time from given a pocket to getting 100 molecules.}).

\section{Proposed Method: \method~}
In this section, we describe the proposed method in detail.
Inspired by the two aforementioned preliminary studies (see Figure \ref{fig:comchange} and Figure \ref{fig:mol_design}), \method~designs drugs for proteins in a two-stage fashion and employs the shape as a bridge since previous work has shown the feasibility of designing drug by molecular shape \cite{shape1,shape2,shape3,shape4,shape5}.

Specifically, \method~designs drugs by first sampling appropriate shapes complementary to the target pocket and then mapping the shapes to specific molecules.
In Section \ref{sec:overview}, we first introduce the overall picture of how \method~works in a zero-shot setting.
Then we pose two challenges: how to sketch the reasonable molecular shapes and how to generate corresponding molecules based on the shapes.
We put forward solutions in Section \ref{sec:sketching} and Section \ref{sec:generate},  respectively.

\subsection{Zero-Shot Pipeline}
\label{sec:overview}

\begin{figure}
     \centering
     \begin{subfigure}[b]{0.32\textwidth}
         \centering
         \includegraphics[width=\textwidth]{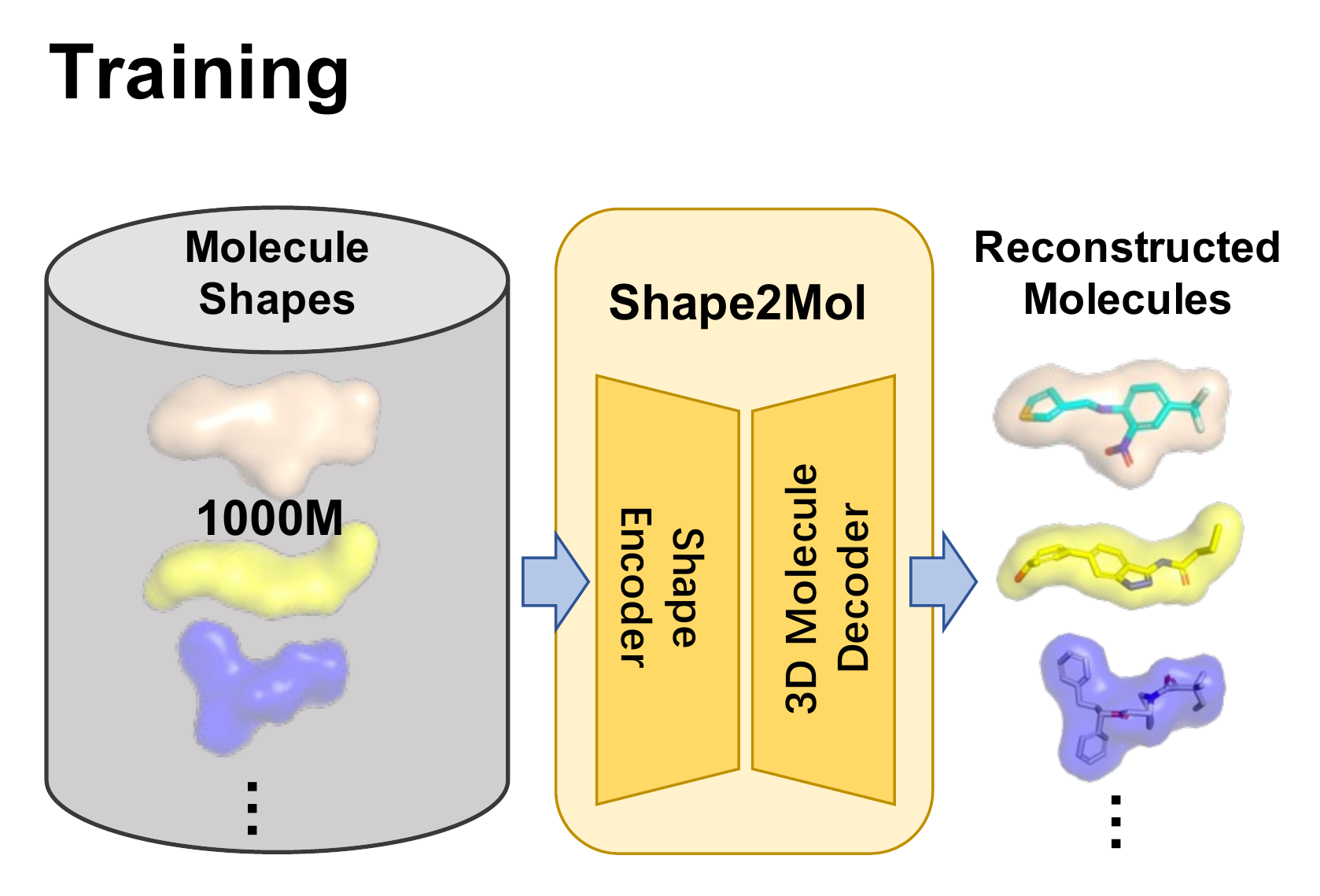}
         \caption{Pre-training with massive unbound molecules}
         \label{fig:training}
     \end{subfigure}\hfill%
     \begin{subfigure}[b]{0.30\textwidth}
         \centering
         \includegraphics[width=\textwidth]{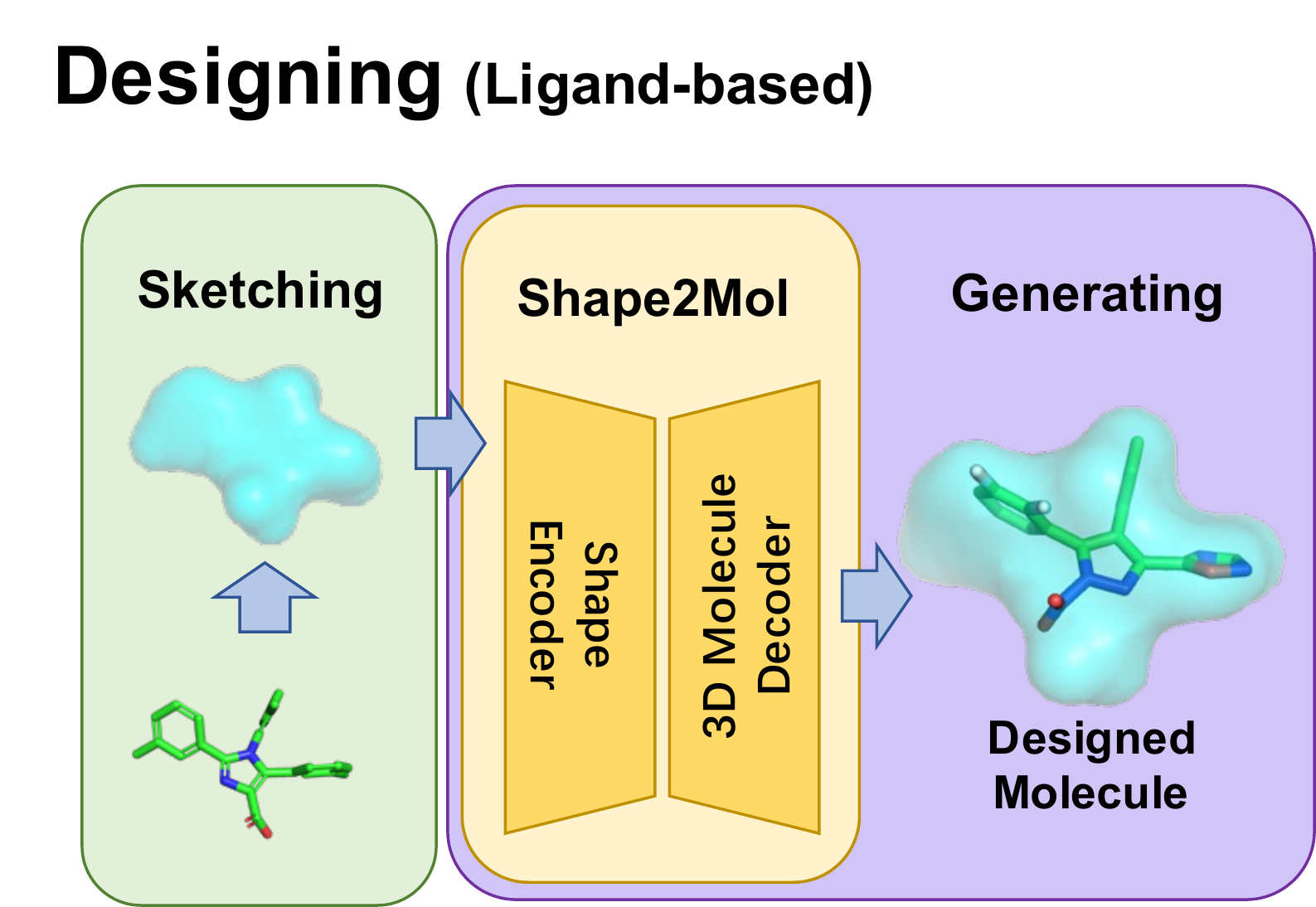}
         \caption{\method~with ligand-based sketching}
         \label{fig:ligand_based}
     \end{subfigure}\hfill
     \begin{subfigure}[b]{0.30\textwidth}
         \centering
         \includegraphics[width=\textwidth]{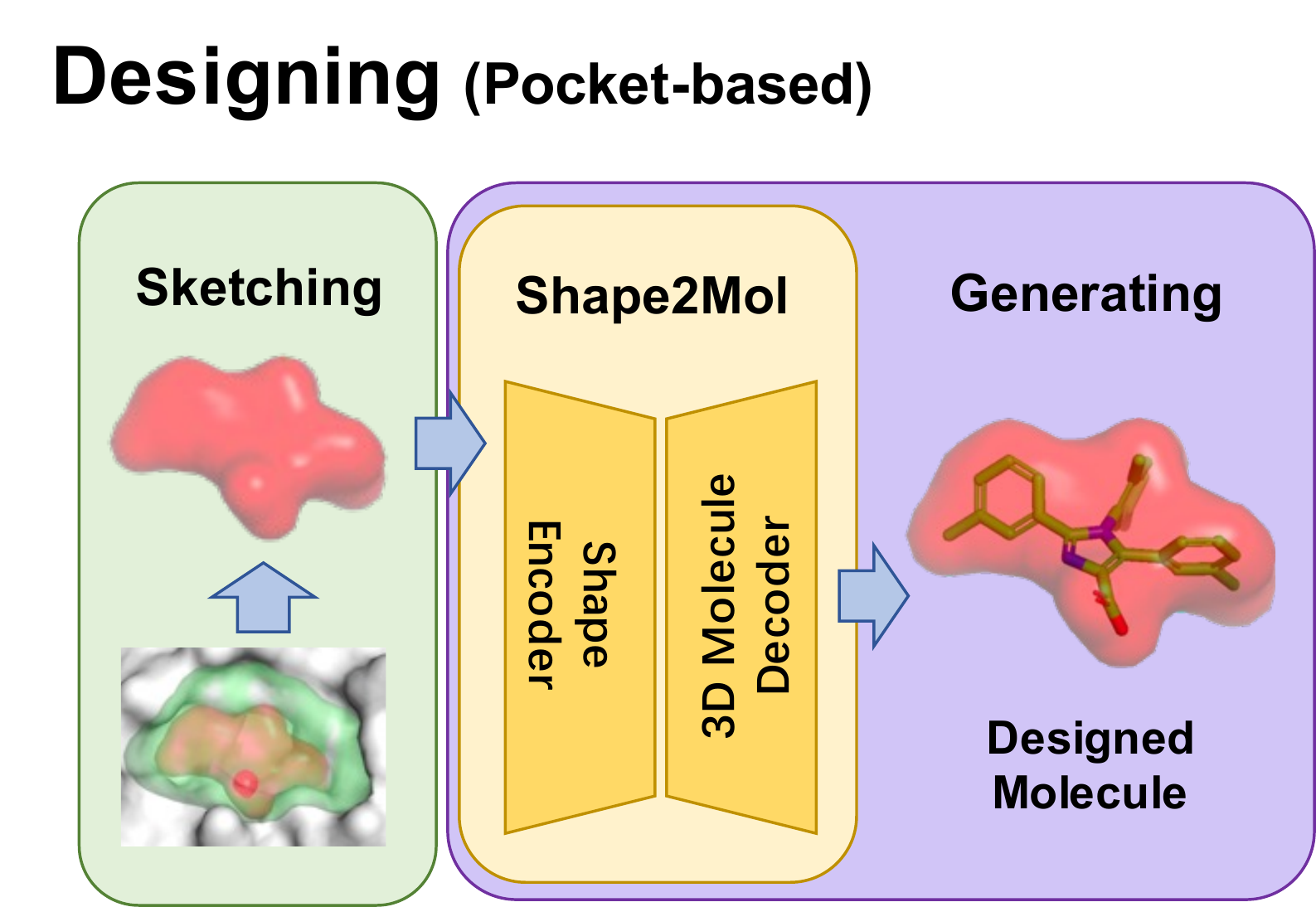}
         \caption{\method~with pocket-based sketching}
         \label{fig:pocket_based}
     \end{subfigure}
        \caption{Overview of \method. In Figure (a), we use massive unbound molecules$^2$ to train the \methodmodel, which includes two sketching variants. In Figure (b) and (c), we use the \methodmodel~to generate 3D molecules to fill in the given shapes. In Figure (b) where existing drugs are available, we treat their shapes as desired molecule shapes (ligand-based). In Figure (c) where only protein pockets are given, we  heuristically sample reasonable molecular shapes from the pockets (pocket-based). }
        \label{fig:overview}
\end{figure}
In this section, we focus on how \method~designs drugs in the zero-shot setting. 
Briefly speaking, \method~produces molecules in a two-stage fashion: sampling the shape of the desired drug first (sketching) and generating molecules conditioned on the resulting shape (generating).

\textbf{Zero-Shot Sketching} 
There are mainly two cases when \method~needs to sample molecular shapes.
In the zero-shot case where no reference protein ligands are available, \method~samples reasonable shapes from protein pockets (see Figure \ref{fig:pocket_based}) based on biological observations.
Besides, \method~can also reuse the shape of a ligand to design a novel one (see Figure \ref{fig:ligand_based}).
Details are listed in Section \ref{sec:sketching}.

\textbf{Zero-Shot Generating} 
\method~generates molecules through a pre-trained generative model, namely \methodmodel, which can convert a given shape into diverse molecules (see Figure \ref{fig:training}).
In this procedure, we utilize massive unbound molecules to train the model. Thus no information about proteins is needed. 
Details of this model are presented in Section \ref{sec:generate}.

\subsection{Sketching Molecular Shapes}
\label{sec:sketching}
The sketching stage is responsible for deciding what desired molecules look like.
In this section, we show how we design two heuristic methods to sketch the shapes of desired molecules.


There are mainly two cases when we sketch a molecule shape based on whether the ligand is provided. 
When the ligand is available, the sketching process can be trivial since molecules with similar shapes have similar properties. 
It is reasonable to directly use the ligand's shape as the shape of the desired molecule, which we call \textbf{Ligand-based Sketching}.
If the ligand of the pocket is unavailable, the challenge is how to obtain a shape that has a high potential to bind to a pocket. \footnote{The pocket can be generated by CAVITY \cite{CAVITY} or f-pocket \cite{Fpocket}. In this study, all protein pockets are generated by CAVITY.}
We name this \textbf{Pocket-based Sketching}, and our main idea is to sample a region with an appropriate size complementary to the surface of the pocket.

Our idea for Pocket-based Sketching is based on two main observations:
\begin{compactenum}
    \item Ligands mainly lie in the area close to the pocket surface. Figure \ref{fig:comchange} shows that the shape of satisfactory ligand is tightly complementary to the pocket.
    \item Pockets are usually much larger than ligands, suggesting that directly utilizing the shape of pockets to design molecules is inappropriate.
\end{compactenum}

\begin{figure}
\centering
\includegraphics[scale=0.5]{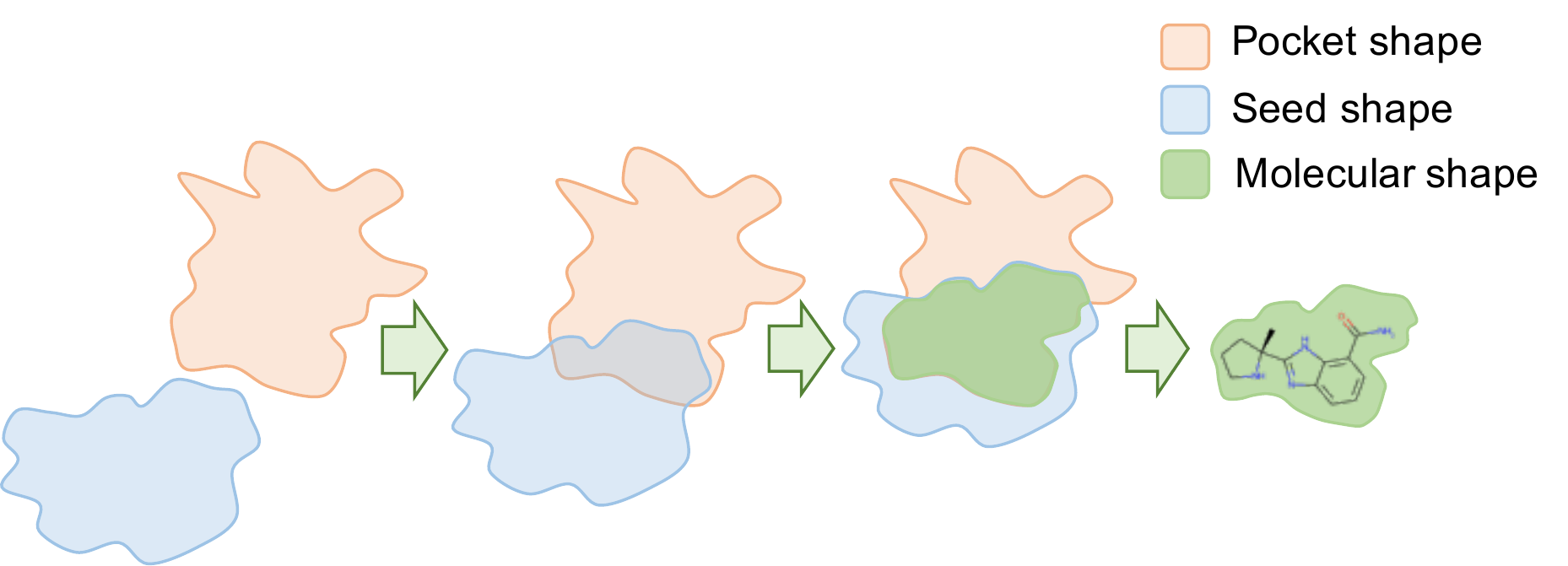}
\caption{A 2D illustration of sampling molecular shapes from pockets. The 2D molecule in the figure represents a potential drug that fits the molecular shape.}
\label{fig:pocket_sampling}
\end{figure}

To this end, we present an algorithm (see the Appendix 1.1)
to obtain the desired molecule shape, which is of the appropriate size and complementary to the pocket surface. We achieve this goal by finding another shape (namely seed shape) that intersects with the pocket, where the intersection has a similar size to a molecule.
We also show a 2D illustration in Figure \ref{fig:pocket_sampling}.


\subsection{Generating 3D Molecules by \methodmodel}
\label{sec:generate}
In this section, we introduce \methodmodel, an encoder-decoder network mapping a shape to diverse and high-quality 3D molecules. There are plenty of unbound molecule data, e.g., 1000M molecules in the ZINC database, making it possible to learn a large-scale pre-trained generative model from shape to molecule.

Concretely,  we formulate the problem as an image-to-sequence generation, where the shape is voxelized as a 3D image (see \ref{para:voxelization}), and the 3D molecule is converted to be a sequence (see \ref{para:linearization}). Our generative approach is capable of modeling any complicated molecule structure and the linearization makes large-scale pre-training easier to implement.



\subsubsection{Encoder: Voxelized Shape}
\label{para:voxelization}



The shape encoder is a 3D extension of the ViT \cite{vit}, where we use 3D patches instead of 2D patches in the original ViT.
Let $\mathcal{A}$ denote the set of all atoms, a molecule $m$ can be constructed as a collection of atoms and their corresponding coordinates:
$$
m=\{ (a, c) | a \in \mathcal{A}, c \in \mathbb{R}^3 \}
$$
Given a molecule $m$, we transform its shape into a 3D image with a voxelization function $v_m: \mathbb{Z}^3 \rightarrow \{0, 1\}$:
$$
v_m(x, y, z) = 
\left\{ \begin{array}{ll}
1 & \exists (a, c) \in m, \|(x, y, z) - c \|_2 \le r(a) + \epsilon\\
0 & \mathrm{otherwise}\\
\end{array} \right.
$$
where $r$ denotes the Van der Waals radii \cite{bondi1964van}, $\epsilon$ is a perturbed noise which helps prevent overfitting \cite{noise}.



\subsubsection{Decoder: Linearized Molecule} 

\label{para:linearization}



The molecule decoder in our model is similar to the Transformer decoder in machine translation \cite{trans}. 
The main difference is that the decoding object here is a 3D molecule instead of a 1D sequence.
To address this, we propose a 3D molecule decoder, which handles a 3D molecule as a sequence of tuples.
The sequence object eases the implementation of a pre-trained model.
To obtain the object, we first cut a molecule into pieces, then convert it to a sequence.  

\textbf{Tokenization} We cut a molecule into pieces so that the generative process can be easily factorized. Our principles are three folds: 
(i) preserving the functional groups since they are vital for determining molecule properties \cite{efg}, 
(ii) avoiding too large size of the vocabulary to ease the pre-training process \cite{xlmr}, 
(iii) no circles exist in the segmented molecules since a tree structure is simpler to handle than a graph.
Our method is simple yet efficient: first tokenizing the molecules with BRCIS \cite{brics} and then
cutting all single bonds attached to a ring. More details about some pilot experiments can be found in Appendix 1.2.


\begin{figure}
    \centering
    \includegraphics[scale=0.25]{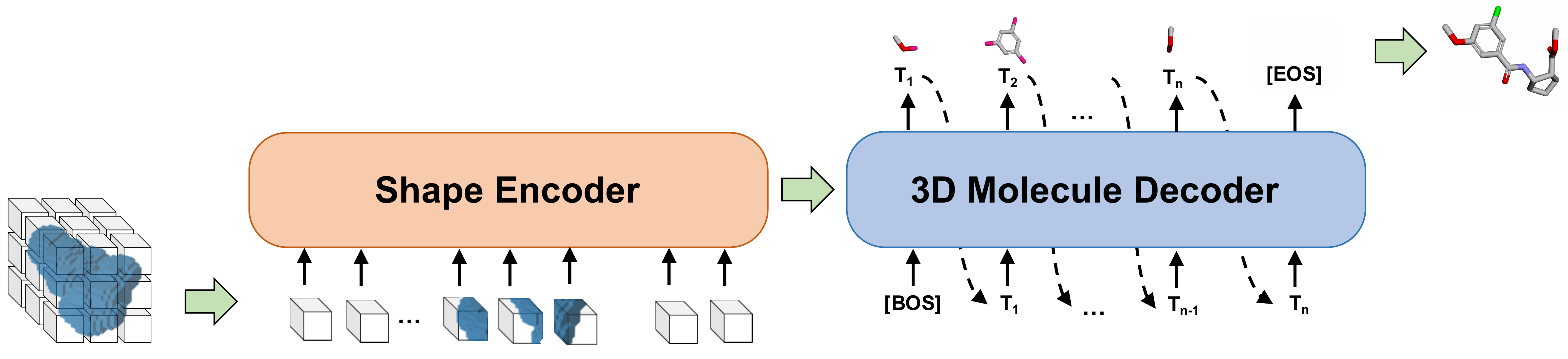}
    \caption{The architecture of \methodmodel.}
    \label{fig:model}
\end{figure}

\begin{figure}
    \centering
    \includegraphics[scale=0.3]{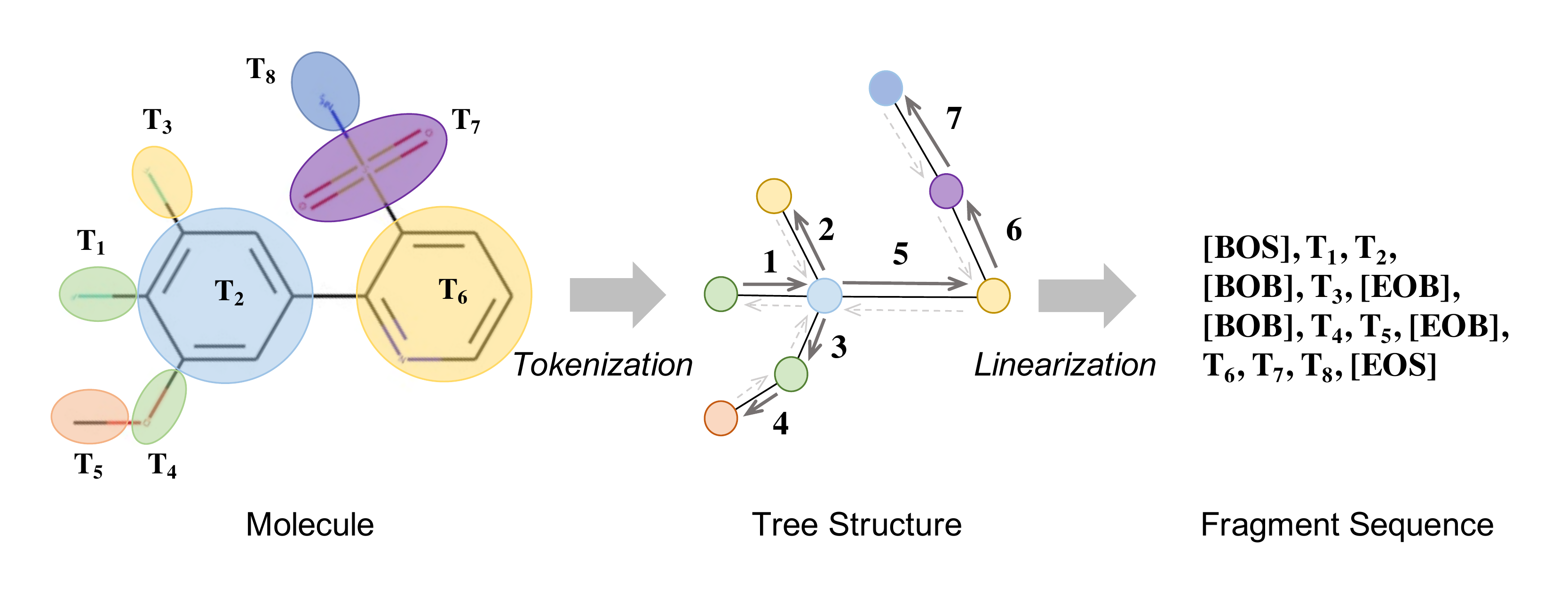}
    \caption{The process of converting a molecule into a sequence.}
    \label{fig:to_sequence}
\end{figure}

\textbf{Linearization} For the network output, we propose to utilize a linearized sequence to represent the target molecule graph, which is not only convenient for training, but also has the strong power to represent any complicated tree structure. 
We first select an fragment whose degree is $1$ as the root of the tree (e.g., \texttt{T1}, \texttt{T3}, \texttt{T5}, and \texttt{T8} in Figure \ref{fig:to_sequence}).
Then we traverse the tree in the depth-first-traverse style\cite{scpn}. Whenever we enter or leave a branch, we will add two special symbols \texttt{[BOB]} and \texttt{[EOB]} (beginning/ending of a branch), respectively.

More particularly, we use a tuple $(C, P, R)$ to represent a fragment $F$, where $C = \mathbf{1}_F$ is an indicator function \cite{wiki_indicator} denoting its index in the vocabulary, $P\in \mathbb{R}^3$ is the translation vector  \cite{wiki_translation}, $R \in \mathbb{R}^4$ is the rotation quaternion \cite{wiki_quaternion}. In order to stabilize the training process, we further discretize the continuous variable $P$ and $R$ into $P^c$ and $R^c$, respectively. Taking the translation vector $P$ as an example, we convert it into a binary vector $P^c$, which satisfies:
$$
P^c[i] = \left\{ \begin{array}{ll}
1 & \lfloor \frac{L}{b}i \rfloor \le P < \lceil \frac{L}{b}i \rceil \\
0 & \mathrm{otherwise}\\
\end{array} \right.
$$

where $L$ is the max translation length,  $b$ is the bin size.

\subsection{Training \& Decoding}
\label{sec:decoding}

\textbf{Training}~~~Given the output probabilities of the model $(\hat{C_{i}}, \hat{P^c_{i}}, \hat{R^c_{i}})$, which denote the probability of a fragment, a discreterized translation vector, and a discreterized rotation vector, respectively. We calculate the corresponding cross-entropy loss and use their sum as the final loss function.
\begin{equation}
    \mathcal{L} = - \sum_{i=1}^{n} \left\{ C_{i}\log \hat{C_{i}} + P^c_{i}\log \hat{P^c_{i}} + R^c_{i} \log \hat{R^c_{i}} \right\}
\end{equation}
where $n$ is the length of fragment sequence.

We use 100M unbound molecules sampled from the lead-like subset of ZINC as the training data.
The Transformer's dimension is 1024, and both encoder and decoder have 12 stacked Transformer layers.
When training \methodmodel, we set the dropout rate as 0.1, batch size 2048, train step 300K and use AdamW \cite{adamw} with learning rate 5e-4, weight decay 1e-2, and warmup step 4000 as the optimizer. The model is developed by ParaGen \footnote{https://github.com/bytedance/ParaGen} and trained on 32 Telsa V100 GPU cards for 2 weeks.
Following \cite{liGAN}, we also randomly rotate and translate the input shape for invariance.

\textbf{Decoding}~~~We design a decoding strategy to provide diverse and high-quality candidate molecules for a given protein.
To achieve diversity, we employ the sampling method Nucleus \cite{topp} to generate multiple fragment sequences.
Then we convert the sequences back to molecules with a greedy algorithm (see Appendix 1.2), which connects the fragments by greedily enumerating the nearest pair of breakpoints. \footnote{Whenever we cut a chemical bond in tokenizing, we mark the atoms of the chemical bond as breakpoints.}
Finally, we do some post-processing operation to further improve the diversity and quality. We remove the duplicate molecules and leverage the docking simulation to drop molecules that do not pass the affinity threshold.

When testing \methodmodel, we set the threshold of Nucleus sampling to 0.95.
For each protein pocket, we sketch 200 shapes.
For each shape, we generate 1000 molecules.
More details of \methodmodel~can be found in the Appendix 1.3.

\section{Results and Discussions}
\subsection{Experiments}
\textbf{Data}~~~
We evaluate the performance of our method on drug design by using a total of 12 proteins (PDB IDs: 1FKG, 2RD6, 3H7W, 3VRJ, 4CG9, 4OQ3, 4PS7, 5E19, 5MKU, 3FI2, 4J71), which is a combination of the test data used in \citetx{liGAN} and \citetx{rational}.
Among the proteins, only JNK3 and GSK3$\beta$ (PDB IDs: 3FI2, 4J71) have sufficient labeled data to train well-performing bioactivity predictors.
We denote the set of these two proteins as Set B and the rest as Set A.
Because 1D/2D methods need bioactivity predictors to design drugs for given proteins, we only evaluate them on Set B while evaluating 3D-based methods on both sets.

\textbf{Baselines}~~~
We compare \method~with some baselines for drug design. Based on the resource needed for designing drugs, we further divide these baselines into three groups:
\textbf{Guided} methods need docking simulation or extra bioactivity predictors to provide supervision signals.
\textbf{Supervised} methods rely on labeled data to train their models.
\textbf{Retrieved} methods directly search the database for desired molecules.

\textbf{Evaluation}~~~
Following \citetx{3dmars}, we evaluate the performance of methods from two aspects. (1) the molecular space covered by designed results. (2) the capacity to provide highly active molecules.
As a high-quality molecular space should contain diverse, novel molecules with pharmaceutical potential, we use five metrics to evaluate the molecular space:
Uniqueness (\textbf{Uniq}), Novelty (\textbf{Nov}), Diversity (\textbf{Div}), Success rate (\textbf{Succ}), and Product (\textbf{Prod}). 
To evaluate the ability to provide highly active molecules, we compare the distributions of $\mathrm{Vina_{score}}$ and use Median Vina Score (\textbf{Median}) to quantify the distribution.
More details about these metrics can be found in Appendix 2.1.

\textbf{Detail of Generation}~~~We describe how different methods generate molecules for comparison.
For each protein, every method needs to generate 100 molecules for comparison.
For 3D methods, we further use the local minimization module in Vina \cite{vina} to optimize the generated structures.
For 1D/2D methods, we first use RDKit to generate the 3D conformer of their results.
For method \textsc{Screen}, we sample 1K and 200K molecules from ZINC and return 100 molecules with the highest $\mathrm{Vina_{score}}$ as the generated results.


\subsection{Main Results}
\label{sec:main_results}
\begin{table}
    \small
    \footnotesize
    \caption{Performance comparison among drug design methods. $\uparrow$ indicates higher is better. $\downarrow$ indicates lower is better.}
    \label{tab:main}
    \begin{center} 
    \begin{tabular}{ c c c c c c c c  c }
    \toprule
    \multirow{2}{*}{Targets} & \multicolumn{2}{c}{\multirow{2}{*}{Method}} & 
    Uniq & Succ & Nov & Div & Prod  & Median\\
    & & &(\%)$\uparrow$ & (\%)$\uparrow$ & (\%)$\uparrow$ & $\uparrow$&$\uparrow$ & (kcal/mol)$\downarrow$ \\ 
    \midrule 
    \multirow{6}{*}{Set A} & Guided & GEKO \cite{3dmars} & 100.0 & 55.7 & 100.0 & 0.912 & 0.51 & -9.58 \\
    \cmidrule(r){2-9}
    & \multirow{2}{*}{Supervised} &liGAN \cite{liGAN} & 100.0 & 0.4 & 100.0 & 0.924 & 0.00 & -5.84 \\
    & &3D SBDD \cite{ShitongLuo} & 69.7 & 13.6 & 98.9 & 0.839 & 0.08 &  -8.83  \\
    \cmidrule(r){2-9}
    & \multirow{2}{*}{Retrieved}& \textsc{Screen} (1K) & 100.0 & 25.6 & 100.0 & 0.892 & 0.23 &  -7.46  \\ 
    & & \textsc{Screen} (200K) & 100.0 & 64.0 & 100.0 & 0.889 & \textbf{0.57} &  -8.66  \\ 
    \cmidrule(r){2-9}
    & \multirow{2}{*}{Ours} & \methodligand~ & 100.0 & 65.3 & 87.0 & 0.786 &  0.41 & -8.89 \\
     &  &\methodpocket~ & 100.0 & 61.1 & 100.0 & 0.908 & \textbf{0.57} & \textbf{-9.62}\\
    \midrule 
    \multirow{13}{*}{Set B} & \multirow{8}{*}{Guided} &JT-VAE \cite{jtvae} & 100.0 & 13.0 & 100.0  & 0.907 &   0.12  &   -8.35 \\
    & &RationaleRL \cite{rational} & 100.0 & 27.0  & 35.0 & 0.884 &   0.08 &   -7.75 \\
    & &GA + D \cite{ga_d} & 39.0 & 24.0 & 87.0 & 0.852 &   0.06  &   -7.22 \\
    & &GraphAF \cite{graphaf} & 97.0 & 0.5 & 100.0 & 0.946 &   0.00  &   -4.22 \\
    & &MolDQN \cite{moldqn} & 76.5 & 0.0 & 100.0 & 0.742 &    0.00  &   -5.52 \\
    & &MolEvol \cite{molevol} & 99.5 & 40.5 & 63.5 & 0.742 &    0.17  &   -8.19\\
    & &MARS \cite{mars} & 86.0 & 31.5 & 93.0 & 0.805 &   0.22 &   -7.68\\
    & & GEKO \cite{3dmars} & 100.0 & 57.0 & 100.0 & 0.910 &  0.52  &  -9.19\\
    \cmidrule(r){2-9}
    & \multirow{2}{*}{Supervised} &liGAN \cite{liGAN} & 99.8 & 0.2 & 100.0 & 0.923 &  0.00  & -5.34\\
    & &3D SBDD \cite{ShitongLuo} & 99.9 & 5.2 & 100.0 & 0.853 &   0.05  &   -6.39\\
    \cmidrule(r){2-9}
    & \multirow{2}{*}{Retrieved}& \textsc{Screen} (1K) & 100.0 & 3.0 & 100.0 & 0.891 & 0.03 &  -6.94  \\ 
    & & \textsc{Screen} (200K) & 100.0 & 32.0 & 100.0 & 0.882 & 0.28 &  -7.95  \\ 
    \cmidrule(r){2-9}
    &\multirow{2}{*}{Ours} &\methodligand~ & 100.0 & 18.0 & 100.0 & 0.913 &   0.17  & -7.34\\
    & &\methodpocket~ & 100.0 & 61.0 & 100.0 & 0.907 &  \textbf{0.55}  & \textbf{-9.32}\\
    \bottomrule
    \end{tabular}
    \end{center}
\end{table}

\begin{figure}
\centering
\begin{minipage}[t]{0.45\textwidth}
\centering
    \includegraphics[scale=0.34]{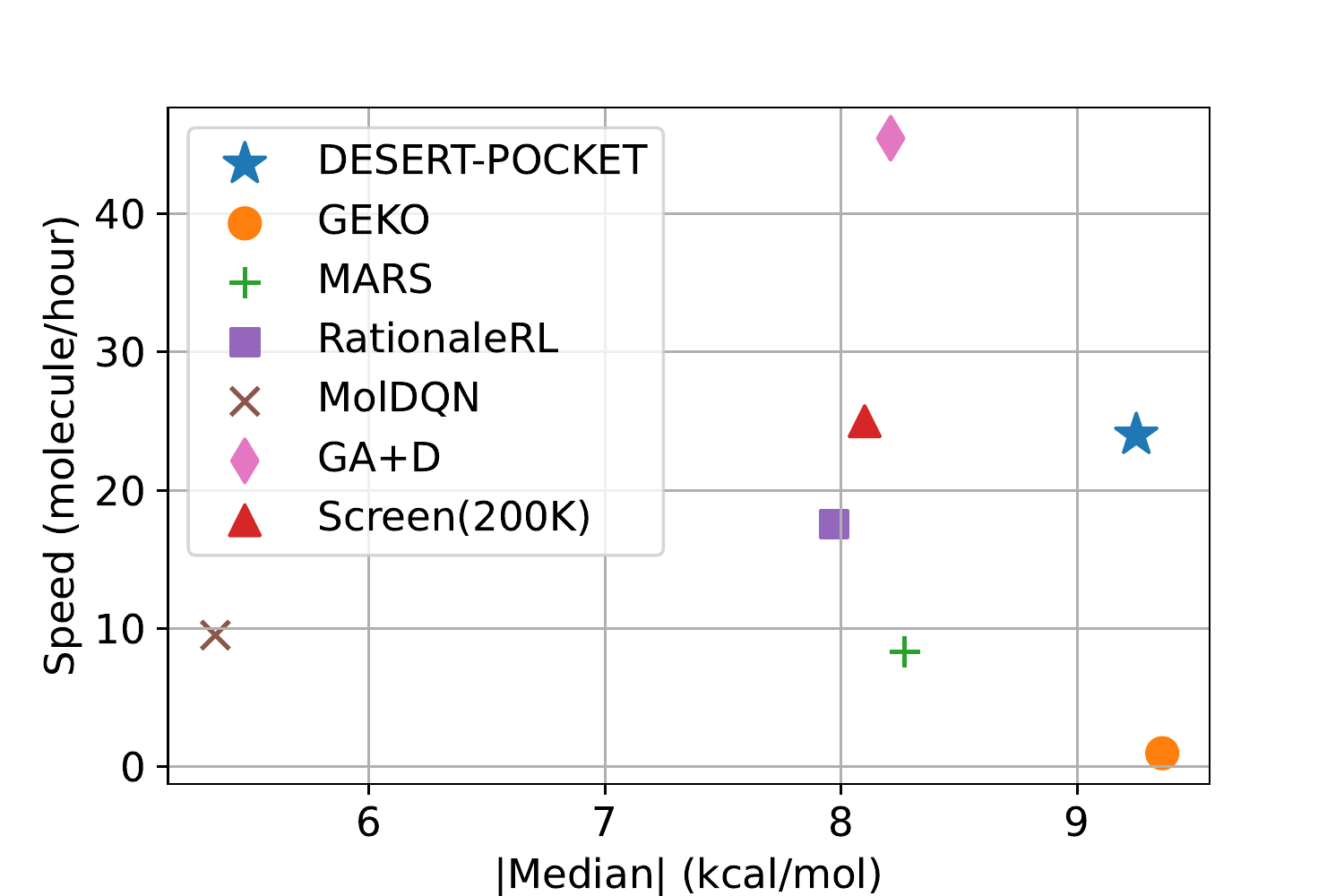}
    \caption{The absolute value of the median vina score and the speed of different methods on target protein 3FI2.}
    \label{fig:speed_median}
\end{minipage}
\hfill
\begin{minipage}[t]{0.45\textwidth}
\centering
\includegraphics[scale=0.6]{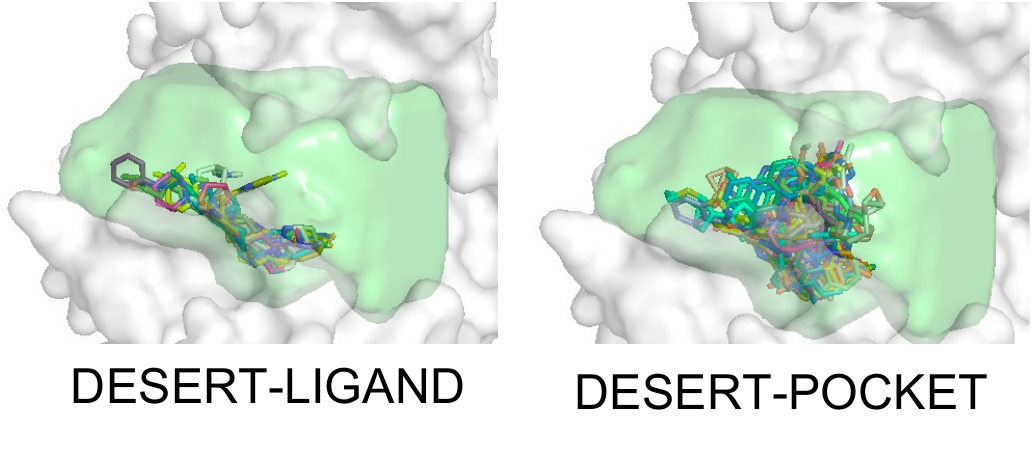}
\caption{Molecules generated by \method{} on protein 4CG9.}
\label{fig:lvsp}
\end{minipage}
\end{figure}

Table \ref{tab:main} shows the results of \method~and baselines. All values are averaged over target proteins.
Our main findings are listed as follows:


\textit{\uppercase\expandafter{\romannumeral1}.The zero-shot \method~achieves the SOTA result at a fast speed.}~~~
\method's performance is strong compared with GEKO, an MCMC-based model with a huge sampling space. 
Compared with GEKO, which works in a trial-and-error way, \method~makes a more clever choice by pruning the space with its biological knowledge regarding the shape and quickly finds a good solution with limited hints from a teacher. We compare the generation speed in Figure \ref{fig:speed_median}.




\textit{\uppercase\expandafter{\romannumeral2}.The shape helps \method~produce high quality molecules.}~~~
The molecular space of \textsc{Screen} is the ZINC database, while that of \method~is generated by the pre-trained model \methodmodel, which is aware of the pocket's shape.
According to the table, both \methodligand~ and \methodpocket~ show observably better performance than their counterparts, i.e., \textsc{Screen} (1K) and \textsc{Screen} (200K).


\textit{\uppercase\expandafter{\romannumeral3}. More comprehensive exploration of protein pockets benefits performance.}~~~
Instead of using the molecular shapes of reference ligands as input, \methodpocket~ comprehensively explores the protein pockets by sampling multiple molecular shapes from them.
Therefore, it has the potential to obtain diverse, high-quality molecules that bind to protein pockets in different regions.
While \methodligand~ only considers one region, i.e., the region that ligands lie in, limiting the exploration of protein pockets.
We show a case in Figure \ref{fig:lvsp}.

\textit{\uppercase\expandafter{\romannumeral4}. Unsupervised methods have larger potential than the supervised counterparts.}~~~
As labeled data is inadequate, e.g., scPDB \footnote{scPDB is a high-quality labeled dataset for 3D drug design.} only has 16,034 entries, the supervised methods easily collapse to the main molecule pattern in their dataset.
When generating molecules with them, we often get the same molecules, e.g., 3D SBDD can only generate 16 unique molecules for protein 4OQ3.
\method~utilizes massive unbound molecules, which leads to the learned space being denser.
Combing with an appropriate sampling method, it can generate diverse molecules.


\subsection{Comparison with Related Shape-based Models}
In this section, we study the shape faithfulness and structure rationality compared with previous shape-based models.
In Table \ref{tab:shape_energy}, we use Shape Tanimoto \footnote{Shape Tanimoto \cite{shape_f1_1,shape_f1_2,shape_f1_3} measures the similarity between the input and generated molecules. $\mathrm{Shape\ Tanimoto}(A, B) = \frac{A\cap B}{A \cup B}$, where $A$ and $B$ are two molecular shapes.
} to evaluate the faithfulness and use Free Energy to quantify the rationality \cite{fenergy}.
We compares \methodligand~ with liGAN (Ligand), a variant of liGAN \cite{liGAN} which utilizes the existing ligands. Although liGAN (Ligand) achieves high performance on Shape Tanimoto, its atom-based decoding strategy does not guarantee the correct relative position between atoms. Therefore, liGAN shows higher Free Energy, indicating that unrealistic structures may appear. Figure \ref{fig:case} shows a case where liGAN produces an ill ring structure.



\begin{table}
	\begin{minipage}{0.44\linewidth}
  \caption{Comparison of shape faithfulness and structure rationality. ``Random'' is the Shape Tanimoto between two random molecules, ``Real'' is the Free Energy of reference ligands.}
  \label{tab:shape_energy}
  \centering
  \resizebox{\textwidth}{!}{
  \begin{tabular}{ccc}
    \toprule
    \multirow{2}{*}{Method} & \multirow{2}{*}{Shape Tanimoto} & Free Energy \\
    & & (kcal/mol)\\
    \midrule 
    Random & 0.325 & / \\
    Real & / & 167.28 \\
    liGAN (Ligand) & 0.869 & 289.55\\
    \methodligand~ & 0.875 & 188.54\\
    \bottomrule
  \end{tabular}
  }
	\end{minipage}
	\hfill
	\begin{minipage}{0.5\linewidth}
		\centering
    \includegraphics[scale=0.6]{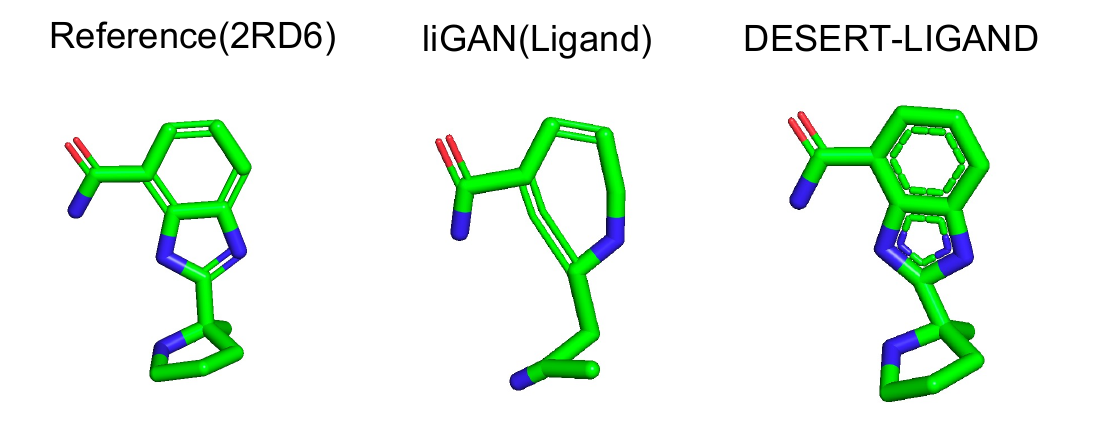}
    \captionof{figure}{A cases from liGAN and \method.} 
    \label{fig:case}
	\end{minipage}
\end{table}

\subsection{Ablation Study of Generating}
In this section, we evaluate several designs of the generating stage in our \method~method, which relate to the pre-trained model and decoding strategy. All results are based on an extra test set from ZINC and a smaller version of \methodmodel. \footnote{The extra test set contains 10K molecules. The smaller \methodmodel~has 512 model dimension and 6 layers of encoder and decoder. We use a greedy decoding strategy and remove post-processing.}

\begin{figure}
\begin{minipage}{0.45\linewidth}
\centering
\includegraphics[scale=0.24]{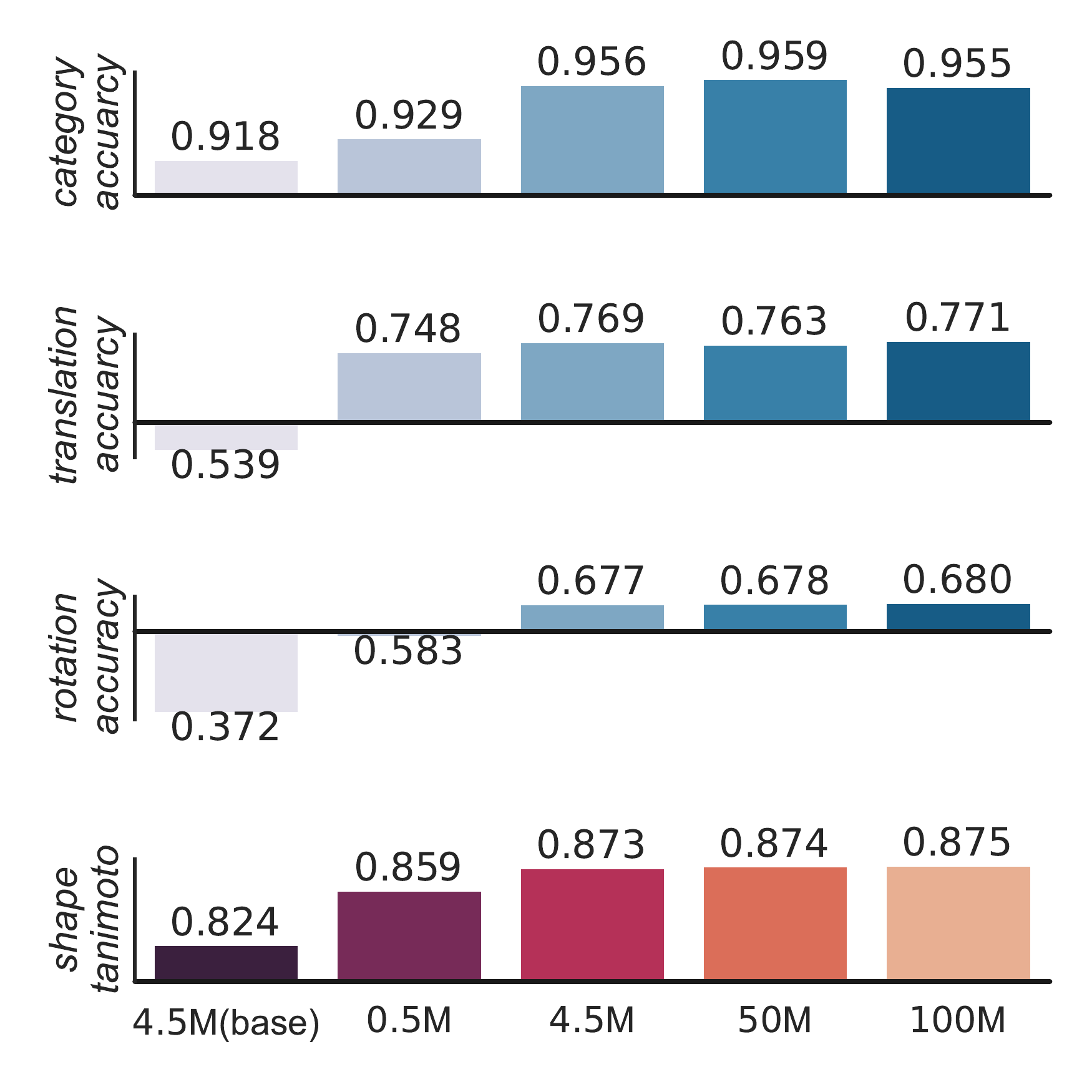}
\caption{Comparison of different pre-training configurations.}
\label{fig:preconfig}
\end{minipage}
\hfill
\begin{minipage}{0.45\linewidth}
\centering
\includegraphics[scale=0.39]{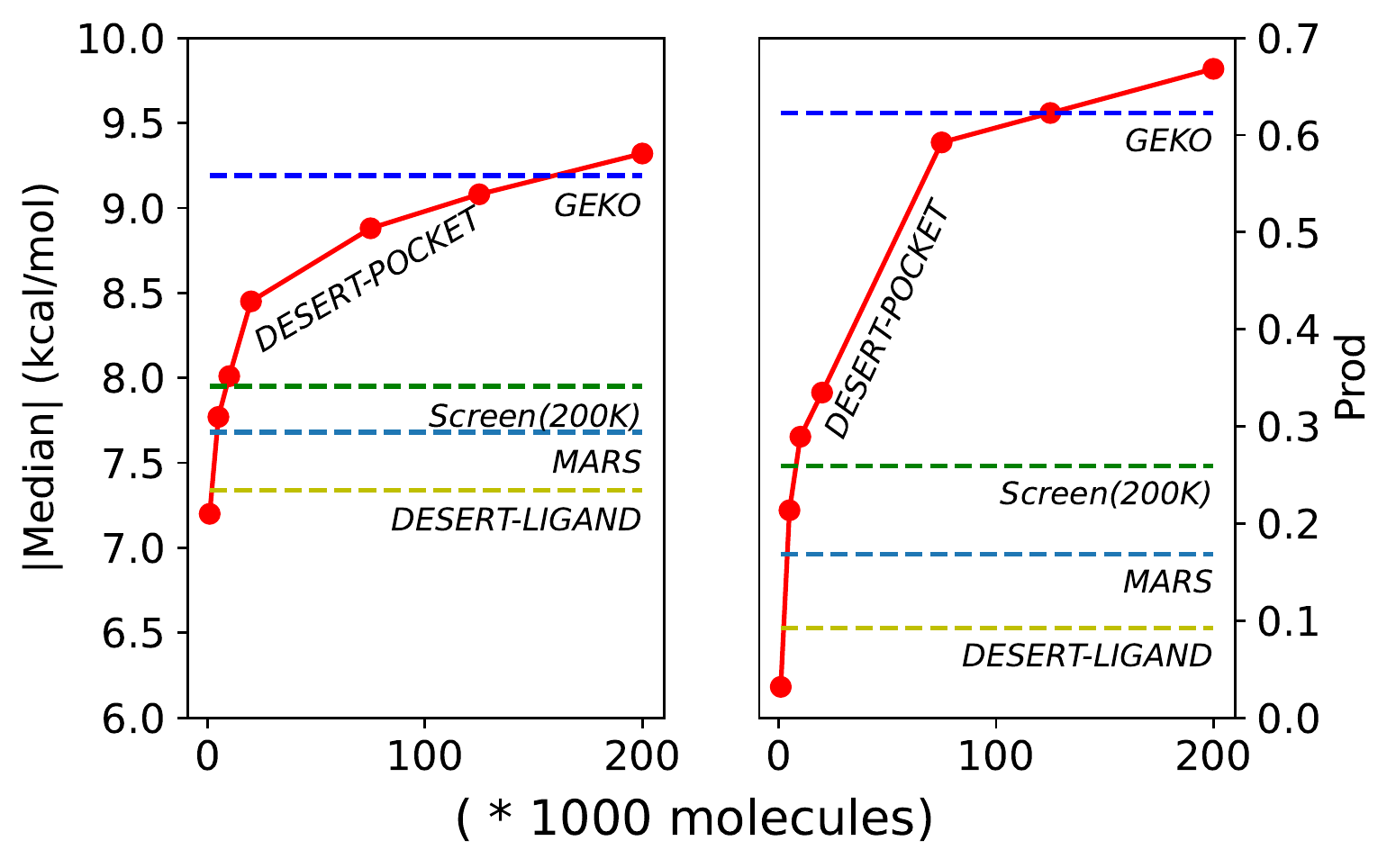}
\captionof{figure}{Comparison of sampling space size.}
\label{fig:sampling_size}
\end{minipage}
\end{figure}

\textbf{Pre-training Configuration}~~~
We evaluate the model quality on different pre-training configurations (mainly focusing on the size of the model and training data).
The results in Figure \ref{fig:preconfig} show: (1) Larger model achieves better performance.
\citetx{bert} observes similar phenomenons in natural language modeling.
(2) Performance saturation occurs when the dataset is of moderate size.
As the map from radii to atoms is easy to learn, the model can capture it with a moderate dataset.
\citetx{roberta} reports a similar result that a large dataset does not necessarily lead to better quality.

\textbf{More Ablation}~~~We also do ablation study about some model variants (including \textit{discretization} and \textit{robust training}) and \textit{decoding strategies}. We further study \textit{chemical information driven design} and \textit{atom-based pre-training}. We refer the reader for more details to the Appendix 2.2.

\begin{table}
\footnotesize
\caption{Performance comparison on more protein targets. Because the original test data only contains incomplete protein pockets, we recover the complete pockets by aligning the incomplete pockets to the structures from PDB database. We managed to recover $55$ pockets and apply \methodpocket~to them. For 3D SBDD, we use the released code and apply the same post-processing used in \method~to it.}
\label{tab:more}
\begin{center}
\begin{tabular}{ c c c c c }
\toprule
\rebuttal{Metric} & \rebuttal{3D SBDD} & \rebuttal{3D SBDD} & \rebuttal{\methodpocket} & \rebuttal{\methodpocket} \\
\rebuttal{Med.} & \rebuttal{w/o post-processing} & \rebuttal{w post-processing} & \rebuttal{w/o post-processing} & \rebuttal{w post-processing} \\
\midrule 
\rebuttal{Vina Score} & \multirow{2}{*}{\rebuttal{-6.069}} & \multirow{2}{*}{\rebuttal{-7.584}} & \multirow{2}{*}{\rebuttal{-6.148}} & \multirow{2}{*}{\rebuttal{-9.410}} \\
\rebuttal{(kcal/mol)} &  &  &  & \\
\midrule
\rebuttal{QED} & \rebuttal{0.522} & \rebuttal{0.501} & \rebuttal{0.614} & \rebuttal{0.549} \\
\midrule
\rebuttal{SA} & \rebuttal{0.672} & \rebuttal{0.623} & \rebuttal{0.612} & \rebuttal{0.616}\\
\midrule
\rebuttal{Diversity} & \rebuttal{0.873} & \rebuttal{0.826} & \rebuttal{0.926} & \rebuttal{0.908}\\
\bottomrule
\end{tabular}
\end{center}
\end{table}

\subsection{Ablation Study of Sketching}
In this section, we study the sketching stage in our \method~method, which includes the effect of sampling space size and seed shape on the method's performance.
All the results are based on the same \methodpocket~method in Section \ref{sec:main_results} and are calculated on Set B.

\textbf{Sampling Space Size}~~~
In Figure \ref{fig:sampling_size}, we evaluate the performance of \method~with respect to sampling space size, i.e., the total number of generated molecules before post-processing.
The results show: (1) Increasing sampling space size leads to better performance. With a larger sampling space, \method~finds more molecular shapes complementary to pockets, leading to a performance rise.
(2) The shape can effectively prune the sampling space for screening. Instead of directly searching molecular space, \method~achieves a similar performance by pruning the space from 200K to 10K with molecular shapes.

\textbf{More Ablation}~~~We also do ablation study about the usage of different \textit{seed shape}. We refer the reader for more details to the Appendix 2.3.





\subsection{Apply \method~to More Protein Targets}
To test the generalization ability of our method more widely, we also apply \method~to the test data from \citetx{ShitongLuo}, which contains $100$ protein targets.
As shown in Table \ref{tab:more}, based on the idea of \textit{structure determines properties}, \methodpocket~generalizes well on different target proteins in both setting, i.e., with/without post-processing.
Supervised methods, like 3D SBDD, hindered by scarce training data, can not generate diverse molecules.
In contrast, training on massive unbound and drug-like molecules, \method~easily generates diverse and promising molecules.
Moreover, sketching molecular shapes based on given pockets also contributes to the better binding affinity of \method.
However, DESERT gives a lower SA score than 3D SBDD.
We assume that it is because the generated molecules of DESERT tend to be structurally complicated, which leads to a slightly worse synthesis score.

\section{Conclusions}
In this study, we propose a zero-shot drug design method \method, which splits the drug design process into two stages: sketching and generating. \method~bridges the two stages with the molecular shape and utilize a large-scale molecular database to reduce the dependence on experimental data and docking simulation. Experiments show that \method~achieves a new state-of-the-art at a fast speed.

\section*{Acknowledgements}
We would like to thank the anonymous reviewers for their insightful comments. Both Hao Zhou and Xinyu Dai are the corresponding authors. This work is jointly supported by Guoqiang Research Institute General Project, Tsinghua University (No. 2021GQG1012) and National Science Foundation of China (No. 61936012 and 61976114).

{
\printbibliography

@article{liGAN,
  author    = {Tomohide Masuda and
               Matthew Ragoza and
               David Ryan Koes},
  title     = {Generating 3D Molecular Structures Conditional on a Receptor Binding
               Site with Deep Generative Models},
  journal   = {CoRR},
  volume    = {abs/2010.14442},
  year      = {2020},
  url       = {https://arxiv.org/abs/2010.14442},
  eprinttype = {arXiv},
  eprint    = {2010.14442},
  bibsource = {dblp computer science bibliography, https://dblp.org}
}

@article{ShitongLuo,
  title={A 3D Generative Model for Structure-Based Drug Design},
  author={Luo, Shitong and Guan, Jiaqi and Ma, Jianzhu and Peng, Jian},
  journal={Advances in Neural Information Processing Systems},
  volume={34},
  year={2021}
}

@misc{
3dmars,
title={Knowledge Guided Geometric Editing for Unsupervised Drug Design},
author={Yuwei Yang and Siqi Ouyang and Meihua Dang and Mingyue Zheng and Lei Li and Hao Zhou},
year={2022},
url={https://openreview.net/forum?id=91muTwt1_t5}
}

@article{bo1,
  title={Automatic chemical design using a data-driven continuous representation of molecules},
  author={G{\'o}mez-Bombarelli, Rafael and Wei, Jennifer N and Duvenaud, David and Hern{\'a}ndez-Lobato, Jos{\'e} Miguel and S{\'a}nchez-Lengeling, Benjam{\'\i}n and Sheberla, Dennis and Aguilera-Iparraguirre, Jorge and Hirzel, Timothy D and Adams, Ryan P and Aspuru-Guzik, Al{\'a}n},
  journal={ACS central science},
  volume={4},
  number={2},
  pages={268--276},
  year={2018},
  publisher={ACS Publications}
}

@inproceedings{jtvae,
  title={Junction tree variational autoencoder for molecular graph generation},
  author={Jin, Wengong and Barzilay, Regina and Jaakkola, Tommi},
  booktitle={International conference on machine learning},
  pages={2323--2332},
  year={2018},
  organization={PMLR}
}

@article{bo2,
  title={Efficient multi-objective molecular optimization in a continuous latent space},
  author={Winter, Robin and Montanari, Floriane and Steffen, Andreas and Briem, Hans and No{\'e}, Frank and Clevert, Djork-Arn{\'e}},
  journal={Chemical science},
  volume={10},
  number={34},
  pages={8016--8024},
  year={2019},
  publisher={Royal Society of Chemistry}
}

@article{molgan,
  title={MolGAN: An implicit generative model for small molecular graphs},
  author={De Cao, Nicola and Kipf, Thomas},
  journal={arXiv preprint arXiv:1805.11973},
  year={2018}
}

@article{rl1,
  title={Deep reinforcement learning for de novo drug design},
  author={Popova, Mariya and Isayev, Olexandr and Tropsha, Alexander},
  journal={Science advances},
  volume={4},
  number={7},
  pages={eaap7885},
  year={2018},
  publisher={American Association for the Advancement of Science}
}

@article{rl2,
  title={Graph convolutional policy network for goal-directed molecular graph generation},
  author={You, Jiaxuan and Liu, Bowen and Ying, Zhitao and Pande, Vijay and Leskovec, Jure},
  journal={Advances in neural information processing systems},
  volume={31},
  year={2018}
}

@article{graphaf,
  title={Graphaf: a flow-based autoregressive model for molecular graph generation},
  author={Shi, Chence and Xu, Minkai and Zhu, Zhaocheng and Zhang, Weinan and Zhang, Ming and Tang, Jian},
  journal={arXiv preprint arXiv:2001.09382},
  year={2020}
}

@article{moldqn,
  title={Optimization of molecules via deep reinforcement learning},
  author={Zhou, Zhenpeng and Kearnes, Steven and Li, Li and Zare, Richard N and Riley, Patrick},
  journal={Scientific reports},
  volume={9},
  number={1},
  pages={1--10},
  year={2019},
  publisher={Nature Publishing Group}
}

@article{ga1,
  title={Guiding deep molecular optimization with genetic exploration},
  author={Ahn, Sungsoo and Kim, Junsu and Lee, Hankook and Shin, Jinwoo},
  journal={Advances in neural information processing systems},
  volume={33},
  pages={12008--12021},
  year={2020}
}

@article{ga_d,
  title={Augmenting genetic algorithms with deep neural networks for exploring the chemical space},
  author={Nigam, AkshatKumar and Friederich, Pascal and Krenn, Mario and Aspuru-Guzik, Al{\'a}n},
  journal={arXiv preprint arXiv:1909.11655},
  year={2019}
}

@article{ga_2,
  title={Evolutionary algorithms for de novo drug design--A survey},
  author={Devi, R Vasundhara and Sathya, S Siva and Coumar, Mohane Selvaraj},
  journal={Applied Soft Computing},
  volume={27},
  pages={543--552},
  year={2015},
  publisher={Elsevier}
}

@article{ga_3,
  title={A graph-based genetic algorithm and generative model/Monte Carlo tree search for the exploration of chemical space},
  author={Jensen, Jan H},
  journal={Chemical science},
  volume={10},
  number={12},
  pages={3567--3572},
  year={2019},
  publisher={Royal Society of Chemistry}
}

@article{sampling1,
  title={Mimosa: Multi-constraint molecule sampling for molecule optimization},
  author={Fu, Tianfan and Xiao, Cao and Li, Xinhao and Glass, Lucas M and Sun, Jimeng},
  journal={arXiv preprint arXiv:2010.02318},
  year={2020}
}

@article{mars,
  title={Mars: Markov molecular sampling for multi-objective drug discovery},
  author={Xie, Yutong and Shi, Chence and Zhou, Hao and Yang, Yuwei and Zhang, Weinan and Yu, Yong and Li, Lei},
  journal={arXiv preprint arXiv:2103.10432},
  year={2021}
}

@inproceedings{rational,
  title={Multi-objective molecule generation using interpretable substructures},
  author={Jin, Wengong and Barzilay, Regina and Jaakkola, Tommi},
  booktitle={International conference on machine learning},
  pages={4849--4859},
  year={2020},
  organization={PMLR}
}

@inproceedings{molevol,
  title={Molecule Optimization by Explainable Evolution},
  author={Chen, Binghong and Wang, Tianzhe and Li, Chengtao and Dai, Hanjun and Song, Le},
  booktitle={International Conference on Learning Representations},
  year={2020}
}

@article{vina,
  title={AutoDock Vina: improving the speed and accuracy of docking with a new scoring function, efficient optimization, and multithreading},
  author={Trott, Oleg and Olson, Arthur J},
  journal={Journal of computational chemistry},
  volume={31},
  number={2},
  pages={455--461},
  year={2010},
  publisher={Wiley Online Library}
}

@article{qed,
  title={Quantifying the chemical beauty of drugs},
  author={Bickerton, G Richard and Paolini, Gaia V and Besnard, J{\'e}r{\'e}my and Muresan, Sorel and Hopkins, Andrew L},
  journal={Nature chemistry},
  volume={4},
  number={2},
  pages={90--98},
  year={2012},
  publisher={Nature Publishing Group}
}

@article{sa,
  title={Estimation of synthetic accessibility score of drug-like molecules based on molecular complexity and fragment contributions},
  author={Ertl, Peter and Schuffenhauer, Ansgar},
  journal={Journal of cheminformatics},
  volume={1},
  number={1},
  pages={1--11},
  year={2009},
  publisher={Springer}
}

@article{brics,
  title={On the Art of Compiling and Using'Drug-Like'Chemical Fragment Spaces},
  author={Degen, J{\"o}rg and Wegscheid-Gerlach, Christof and Zaliani, Andrea and Rarey, Matthias},
  journal={ChemMedChem: Chemistry Enabling Drug Discovery},
  volume={3},
  number={10},
  pages={1503--1507},
  year={2008},
  publisher={Wiley Online Library}
}

@article{trans,
  title={Attention is all you need},
  author={Vaswani, Ashish and Shazeer, Noam and Parmar, Niki and Uszkoreit, Jakob and Jones, Llion and Gomez, Aidan N and Kaiser, {\L}ukasz and Polosukhin, Illia},
  journal={Advances in neural information processing systems},
  volume={30},
  year={2017}
}

@article{vit,
  title={An image is worth 16x16 words: Transformers for image recognition at scale},
  author={Dosovitskiy, Alexey and Beyer, Lucas and Kolesnikov, Alexander and Weissenborn, Dirk and Zhai, Xiaohua and Unterthiner, Thomas and Dehghani, Mostafa and Minderer, Matthias and Heigold, Georg and Gelly, Sylvain and others},
  journal={arXiv preprint arXiv:2010.11929},
  year={2020}
}

@article{topp,
  title={The curious case of neural text degeneration},
  author={Holtzman, Ari and Buys, Jan and Du, Li and Forbes, Maxwell and Choi, Yejin},
  journal={arXiv preprint arXiv:1904.09751},
  year={2019}
}

@article{Fpocket,
  title={Fpocket: an open source platform for ligand pocket detection},
  author={Le Guilloux, Vincent and Schmidtke, Peter and Tuffery, Pierre},
  journal={BMC bioinformatics},
  volume={10},
  number={1},
  pages={1--11},
  year={2009},
  publisher={BioMed Central}
}

@article{CAVITY,
  title={Binding site detection and druggability prediction of protein targets for structure-based drug design},
  author={Yuan, Yaxia and Pei, Jianfeng and Lai, Luhua},
  journal={Current pharmaceutical design},
  volume={19},
  number={12},
  pages={2326--2333},
  year={2013},
  publisher={Bentham Science Publishers}
}

@article{screen,
  title={Integration of virtual and high-throughput screening},
  author={Bajorath, J{\"u}rgen},
  journal={Nature Reviews Drug Discovery},
  volume={1},
  number={11},
  pages={882--894},
  year={2002},
  publisher={Nature Publishing Group}
}

@article{docking1,
  title={Molecular docking and structure-based drug design strategies},
  author={Ferreira, Leonardo G and Dos Santos, Ricardo N and Oliva, Glaucius and Andricopulo, Adriano D},
  journal={Molecules},
  volume={20},
  number={7},
  pages={13384--13421},
  year={2015},
  publisher={Multidisciplinary Digital Publishing Institute}
}

@article{docking2,
  title={Molecular docking: a powerful approach for structure-based drug discovery},
  author={Meng, Xuan-Yu and Zhang, Hong-Xing and Mezei, Mihaly and Cui, Meng},
  journal={Current computer-aided drug design},
  volume={7},
  number={2},
  pages={146--157},
  year={2011},
  publisher={Bentham Science Publishers}
}

@article{docking3,
  title={Protein-ligand blind docking using QuickVina-W with inter-process spatio-temporal integration},
  author={Hassan, Nafisa M and Alhossary, Amr A and Mu, Yuguang and Kwoh, Chee-Keong},
  journal={Scientific reports},
  volume={7},
  number={1},
  pages={1--13},
  year={2017},
  publisher={Nature Publishing Group}
}

@article{string1,
  title={Conditional molecular design with deep generative models},
  author={Kang, Seokho and Cho, Kyunghyun},
  journal={Journal of chemical information and modeling},
  volume={59},
  number={1},
  pages={43--52},
  year={2018},
  publisher={ACS Publications}
}

@article{string2,
  title={Generating focused molecule libraries for drug discovery with recurrent neural networks},
  author={Segler, Marwin HS and Kogej, Thierry and Tyrchan, Christian and Waller, Mark P},
  journal={ACS central science},
  volume={4},
  number={1},
  pages={120--131},
  year={2018},
  publisher={ACS Publications}
}

@inproceedings{string3,
  title={Grammar variational autoencoder},
  author={Kusner, Matt J and Paige, Brooks and Hern{\'a}ndez-Lobato, Jos{\'e} Miguel},
  booktitle={International conference on machine learning},
  pages={1945--1954},
  year={2017},
  organization={PMLR}
}

@article{graph1,
  title={Constrained graph variational autoencoders for molecule design},
  author={Liu, Qi and Allamanis, Miltiadis and Brockschmidt, Marc and Gaunt, Alexander},
  journal={Advances in neural information processing systems},
  volume={31},
  year={2018}
}

@article{graph2,
  title={Constrained generation of semantically valid graphs via regularizing variational autoencoders},
  author={Ma, Tengfei and Chen, Jie and Xiao, Cao},
  journal={Advances in Neural Information Processing Systems},
  volume={31},
  year={2018}
}

@article{graph3,
  title={Nevae: A deep generative model for molecular graphs},
  author={Samanta, Bidisha and De, Abir and Jana, Gourhari and G{\'o}mez, Vicen{\c{c}} and Chattaraj, Pratim Kumar and Ganguly, Niloy and Gomez-Rodriguez, Manuel},
  journal={Journal of machine learning research. 2020 Apr; 21 (114): 1-33},
  year={2020},
  publisher={Journal of Machine Learning Research}
}

@article{graph4,
  title={Discrete object generation with reversible inductive construction},
  author={Seff, Ari and Zhou, Wenda and Damani, Farhan and Doyle, Abigail and Adams, Ryan P},
  journal={Advances in Neural Information Processing Systems},
  volume={32},
  year={2019}
}

@article{bio,
  title={Composing molecules with multiple property constraints},
  author={Jin, Wengong and Barzilay, Regina and Jaakkola, Tommi},
  journal={arXiv preprint arXiv:2002.03244},
  year={2020}
}

@article{shape1,
  title={Shape-based generative modeling for de novo drug design},
  author={Skalic, Miha and Jim{\'e}nez, Jos{\'e} and Sabbadin, Davide and De Fabritiis, Gianni},
  journal={Journal of chemical information and modeling},
  volume={59},
  number={3},
  pages={1205--1214},
  year={2019},
  publisher={ACS Publications}
}

@article{shape2,
  title={Advances in the development of shape similarity methods and their application in drug discovery},
  author={Kumar, Ashutosh and Zhang, Kam YJ},
  journal={Frontiers in chemistry},
  volume={6},
  pages={315},
  year={2018},
  publisher={Frontiers}
}

@article{shape3,
  title={Drug screening using shape-based virtual screening and in vitro experimental models of cutaneous Leishmaniasis},
  author={Santos, Camila Cardoso and Batista, Marcos Meuser and Ullah, Asma Inam and Reddy, Tummala Rama Krishna and Soeiro, Maria de Nazar{\'e} Correia},
  journal={Parasitology},
  volume={148},
  number={1},
  pages={98--104},
  year={2021},
  publisher={Cambridge University Press}
}

@article{shape4,
  title={A comprehensive survey of small-molecule binding pockets in proteins},
  author={Gao, Mu and Skolnick, Jeffrey},
  journal={PLoS computational biology},
  volume={9},
  number={10},
  pages={e1003302},
  year={2013},
  publisher={Public Library of Science San Francisco, USA}
}

@article{shape5,
  title={Structure-based drug design: aiming for a perfect fit},
  author={Van Montfort, Rob LM and Workman, Paul},
  journal={Essays in biochemistry},
  volume={61},
  number={5},
  pages={431--437},
  year={2017},
  publisher={Portland Press Ltd.}
}

@article{conf1,
  title={Learning neural generative dynamics for molecular conformation generation},
  author={Xu, Minkai and Luo, Shitong and Bengio, Yoshua and Peng, Jian and Tang, Jian},
  journal={arXiv preprint arXiv:2102.10240},
  year={2021}
}

@article{conf2,
  title={Better informed distance geometry: using what we know to improve conformation generation},
  author={Riniker, Sereina and Landrum, Gregory A},
  journal={Journal of chemical information and modeling},
  volume={55},
  number={12},
  pages={2562--2574},
  year={2015},
  publisher={ACS Publications}
}

@article{adamw,
  title={Decoupled weight decay regularization},
  author={Loshchilov, Ilya and Hutter, Frank},
  journal={arXiv preprint arXiv:1711.05101},
  year={2017}
}

@article{shape_f1_1,
  title={Shape-based virtual screening with volumetric aligned molecular shapes},
  author={Koes, David Ryan and Camacho, Carlos J},
  journal={Journal of computational chemistry},
  volume={35},
  number={25},
  pages={1824--1834},
  year={2014},
  publisher={Wiley Online Library}
}

@article{shape_f1_2,
  title={A shape-based 3-D scaffold hopping method and its application to a bacterial protein- protein interaction},
  author={Rush, Thomas S and Grant, J Andrew and Mosyak, Lidia and Nicholls, Anthony},
  journal={Journal of medicinal chemistry},
  volume={48},
  number={5},
  pages={1489--1495},
  year={2005},
  publisher={ACS Publications}
}

@article{shape_f1_3,
  title={PubChem3D: shape compatibility filtering using molecular shape quadrupoles},
  author={Kim, Sunghwan and Bolton, Evan E and Bryant, Stephen H},
  journal={Journal of cheminformatics},
  volume={3},
  number={1},
  pages={1--14},
  year={2011},
  publisher={Springer}
}

@article{efg,
  title={Extended functional groups (EFG): an efficient set for chemical characterization and structure-activity relationship studies of chemical compounds},
  author={Salmina, Elena S and Haider, Norbert and Tetko, Igor V},
  journal={Molecules},
  volume={21},
  number={1},
  pages={1},
  year={2015},
  publisher={MDPI}
}

@article{xlmr,
  title={Unsupervised cross-lingual representation learning at scale},
  author={Conneau, Alexis and Khandelwal, Kartikay and Goyal, Naman and Chaudhary, Vishrav and Wenzek, Guillaume and Guzm{\'a}n, Francisco and Grave, Edouard and Ott, Myle and Zettlemoyer, Luke and Stoyanov, Veselin},
  journal={arXiv preprint arXiv:1911.02116},
  year={2019}
}

@article{bound-notation,
  title={When is it important to measure unbound drug in evaluating nanomedicine pharmacokinetics?},
  author={Stern, Stephan T and Martinez, Marilyn N and Stevens, David M},
  journal={Drug Metabolism and Disposition},
  volume={44},
  number={12},
  pages={1934--1939},
  year={2016},
  publisher={ASPET}
}

@article{mss,
  title={Structure-aware generation of drug-like molecules},
  author={Drot{\'a}r, Pavol and Jamasb, Arian Rokkum and Day, Ben and Cangea, C{\u{a}}t{\u{a}}lina and Li{\`o}, Pietro},
  journal={arXiv preprint arXiv:2111.04107},
  year={2021}
}

@article{bert,
  title={Bert: Pre-training of deep bidirectional transformers for language understanding},
  author={Devlin, Jacob and Chang, Ming-Wei and Lee, Kenton and Toutanova, Kristina},
  journal={arXiv preprint arXiv:1810.04805},
  year={2018}
}

@article{roberta,
  title={Roberta: A robustly optimized bert pretraining approach},
  author={Liu, Yinhan and Ott, Myle and Goyal, Naman and Du, Jingfei and Joshi, Mandar and Chen, Danqi and Levy, Omer and Lewis, Mike and Zettlemoyer, Luke and Stoyanov, Veselin},
  journal={arXiv preprint arXiv:1907.11692},
  year={2019}
}

@inproceedings{noise,
  title={Improving the robustness of deep neural networks via stability training},
  author={Zheng, Stephan and Song, Yang and Leung, Thomas and Goodfellow, Ian},
  booktitle={Proceedings of the ieee conference on computer vision and pattern recognition},
  pages={4480--4488},
  year={2016}
}

@article{pocket,
  title={Pocket-based drug design: exploring pocket space},
  author={Zheng, Xiliang and Gan, LinFeng and Wang, Erkang and Wang, Jin},
  journal={The AAPS journal},
  volume={15},
  number={1},
  pages={228--241},
  year={2013},
  publisher={Springer}
}

@article{md1,
  title={Role of molecular dynamics and related methods in drug discovery},
  author={De Vivo, Marco and Masetti, Matteo and Bottegoni, Giovanni and Cavalli, Andrea},
  journal={Journal of medicinal chemistry},
  volume={59},
  number={9},
  pages={4035--4061},
  year={2016},
  publisher={ACS Publications}
}

@article{md2,
  title={Molecular dynamics simulations and drug discovery},
  author={Durrant, Jacob D and McCammon, J Andrew},
  journal={BMC biology},
  volume={9},
  number={1},
  pages={1--9},
  year={2011},
  publisher={BioMed Central}
}

@article{sdp1,
  title={Computational Chemistry Approaches for Understanding how Structure Determines Properties},
  author={Katritzky, Alan R and Slavov, Svetoslav and Radzvilovits, Maksim and Stoyanova-Slavova, Iva and Karelson, Mati},
  journal={Zeitschrift f{\"u}r Naturforschung B},
  volume={64},
  number={6},
  pages={773--777},
  year={2009},
  publisher={Verlag der Zeitschrift f{\"u}r Naturforschung}
}

@techreport{sdp2,
  title={How Structure Determines Properties of Plastics},
  author={Kell, RM and Stickney, PB},
  year={1964},
  institution={BATTELLE MEMORIAL INST COLUMBUS OHIO}
}

@article{sdp3,
  title={How chemical structure determines physical, chemical, and technological properties: An overview illustrating the potential of quantitative structure- property relationships for fuels science},
  author={Katritzky, Alan R and Fara, Dan C},
  journal={Energy \& Fuels},
  volume={19},
  number={3},
  pages={922--935},
  year={2005},
  publisher={ACS Publications}
}

@article{sdp4,
  title={Ligand Structure Determines Nanoparticles' Atomic Structure, Metal-Ligand Interface and Properties},
  author={Rambukwella, Milan and Sakthivel, Naga Arjun and Delcamp, Jared H and Sementa, Luca and Fortunelli, Alessandro and Dass, Amala},
  journal={Frontiers in chemistry},
  pages={330},
  year={2018},
  publisher={Frontiers}
}

@article{bondi1964van,
  title={van der Waals volumes and radii},
  author={Bondi, A van},
  journal={The Journal of physical chemistry},
  volume={68},
  number={3},
  pages={441--451},
  year={1964},
  publisher={ACS Publications}
}

@misc{wiki_indicator, 
title={Indicator function}, url={https://en.wikipedia.org/wiki/Indicator_function}, 
journal={Wikipedia}, 
publisher={Wikimedia Foundation}
}

@misc{wiki_translation, 
title={Translation (geometry)}, url={https://en.wikipedia.org/wiki/Translation_(geometry)}, 
journal={Wikipedia}, 
publisher={Wikimedia Foundation}
}

@misc{wiki_quaternion, 
title={Quaternion}, 
url={https://en.wikipedia.org/wiki/Quaternion}, 
journal={Wikipedia}, 
publisher={Wikimedia Foundation}
}

@article{dockreliable,
  title={Is it reliable to use common molecular docking methods for comparing the binding affinities of enantiomer pairs for their protein target?},
  author={Ram{\'\i}rez, David and Caballero, Julio},
  journal={International journal of molecular sciences},
  volume={17},
  number={4},
  pages={525},
  year={2016},
  publisher={Multidisciplinary Digital Publishing Institute}
}

@article{dockreliable1,
  title={Combining machine learning systems and multiple docking simulation packages to improve docking prediction reliability for network pharmacology},
  author={Hsin, Kun-Yi and Ghosh, Samik and Kitano, Hiroaki},
  journal={PloS one},
  volume={8},
  number={12},
  pages={e83922},
  year={2013},
  publisher={Public Library of Science San Francisco, USA}
}

@article{docktime,
  title={Lean-Docking: Exploiting Ligands’ Predicted Docking Scores to Accelerate Molecular Docking},
  author={Berenger, Francois and Kumar, Ashutosh and Zhang, Kam YJ and Yamanishi, Yoshihiro},
  journal={Journal of Chemical Information and Modeling},
  volume={61},
  number={5},
  pages={2341--2352},
  year={2021},
  publisher={ACS Publications}
}

@article{docktime1,
  title={CaverDock: a molecular docking-based tool to analyse ligand transport through protein tunnels and channels},
  author={Vavra, Ondrej and Filipovic, Jiri and Plhak, Jan and Bednar, David and Marques, Sergio M and Brezovsky, Jan and Stourac, Jan and Matyska, Ludek and Damborsky, Jiri},
  journal={Bioinformatics},
  volume={35},
  number={23},
  pages={4986--4993},
  year={2019},
  publisher={Oxford University Press}
}

@article{graphmap,
  title={Learning multimodal graph-to-graph translation for molecular optimization},
  author={Jin, Wengong and Yang, Kevin and Barzilay, Regina and Jaakkola, Tommi},
  journal={arXiv preprint arXiv:1812.01070},
  year={2018}
}

@article{graphga,
  title={A graph-based genetic algorithm and generative model/Monte Carlo tree search for the exploration of chemical space},
  author={Jensen, Jan H},
  journal={Chemical science},
  volume={10},
  number={12},
  pages={3567--3572},
  year={2019},
  publisher={Royal Society of Chemistry}
}

@inproceedings{scpn,
    title = "Adversarial Example Generation with Syntactically Controlled Paraphrase Networks",
    author = "Iyyer, Mohit  and Wieting, John  and Gimpel, Kevin  and Zettlemoyer, Luke",
    booktitle = "Proceedings of the 2018 Conference of the North {A}merican Chapter of the Association for Computational Linguistics: Human Language Technologies, Volume 1 (Long Papers)",
    month = jun,
    year = "2018",
    address = "New Orleans, Louisiana",
    publisher = "Association for Computational Linguistics",
    url = "https://aclanthology.org/N18-1170",
    doi = "10.18653/v1/N18-1170",
    pages = "1875--1885",
}

@article{vina2,
  title={AutoDock Vina 1.2. 0: New docking methods, expanded force field, and python bindings},
  author={Eberhardt, Jerome and Santos-Martins, Diogo and Tillack, Andreas F and Forli, Stefano},
  journal={Journal of Chemical Information and Modeling},
  volume={61},
  number={8},
  pages={3891--3898},
  year={2021},
  publisher={ACS Publications}
}

@article{fenergy,
  title={A realistic molecular model of cement hydrates},
  author={Pellenq, Roland J-M and Kushima, Akihiro and Shahsavari, Rouzbeh and Van Vliet, Krystyn J and Buehler, Markus J and Yip, Sidney and Ulm, Franz-Josef},
  journal={Proceedings of the National Academy of Sciences},
  volume={106},
  number={38},
  pages={16102--16107},
  year={2009},
  publisher={National Acad Sciences}
}
}

\section*{Checklist}

\begin{enumerate}

\item For all authors...
\begin{enumerate}
  \item Do the main claims made in the abstract and introduction accurately reflect the paper's contributions and scope?
    \answerYes{}
  \item Did you describe the limitations of your work?
    \answerYes{}
  \item Did you discuss any potential negative societal impacts of your work?
    \answerNo{}
  \item Have you read the ethics review guidelines and ensured that your paper conforms to them?
    \answerYes{}
\end{enumerate}

\item If you are including theoretical results...
\begin{enumerate}
  \item Did you state the full set of assumptions of all theoretical results?
    \answerNA{}
        \item Did you include complete proofs of all theoretical results?
    \answerNA{}
\end{enumerate}

\item If you ran experiments...
\begin{enumerate}
  \item Did you include the code, data, and instructions needed to reproduce the main experimental results (either in the supplemental material or as a URL)?
    \answerYes{}
  \item Did you specify all the training details (e.g., data splits, hyperparameters, how they were chosen)?
    \answerYes{}
        \item Did you report error bars (e.g., with respect to the random seed after running experiments multiple times)?
    \answerNo{}
        \item Did you include the total amount of compute and the type of resources used (e.g., type of GPUs, internal cluster, or cloud provider)?
    \answerYes{}
\end{enumerate}

\item If you are using existing assets (e.g., code, data, models) or curating/releasing new assets...
\begin{enumerate}
  \item If your work uses existing assets, did you cite the creators?
    \answerYes{}
  \item Did you mention the license of the assets?
    \answerNo{}
  \item Did you include any new assets either in the supplemental material or as a URL?
    \answerNo{}
  \item Did you discuss whether and how consent was obtained from people whose data you're using/curating?
    \answerYes{}
  \item Did you discuss whether the data you are using/curating contains personally identifiable information or offensive content?
    \answerNo{}
\end{enumerate}

\item If you used crowdsourcing or conducted research with human subjects...
\begin{enumerate}
  \item Did you include the full text of instructions given to participants and screenshots, if applicable?
    \answerNA{}
  \item Did you describe any potential participant risks, with links to Institutional Review Board (IRB) approvals, if applicable?
    \answerNA{}
  \item Did you include the estimated hourly wage paid to participants and the total amount spent on participant compensation?
    \answerNA{}
\end{enumerate}

\end{enumerate}




\section*{Supplementary material}

# Appendix

## 1 Method

### 1.1 Sketching

In Pocket-based Sketching, we present an algorithm to obtain the molecule shape, which is of the appropriate size and complementary to protein pockets. The algorithm uses a seed shape to intersect with the pocket gradually.

To obtain the seed shape, we first randomly sample several molecules from ZINC, then use the overlapping of their shapes as the seed shape. By using the overlapping strategy, we make the sketched pseudo molecular shapes more native-molecule-like and not overly dependent on one specific molecule. Algorithm 1 is the pseudo-code.

We set the volume threshold to $300\text{Å}^3$, the average volume of molecules. The step size is the same as the voxel resolution, i.e., $0.5\text{Å}$. We initially place the seed shape in a random position as long as the seed shape and the pocket shape do not overlap.

### 1.2 Generating

**Pilot Experiments of Tokenizing Methods** We evaluate different molecule tokenizing methods based on three principles (preserving functional groups, appropriate vocabulary size, and no circles structures) and propose our approach in the context of pre-training. There are several popular tokenizing methods for drug design: Atom-based [1, 2] shatters a molecule into atoms. Cut Single Bond [3] cuts all single bonds in a molecule. Both RECAP [4] and BRICS [5] use chemical reaction rules for preserving functional groups. The results in Table 1 show: (1) Atom-based and Cut Single Bond can not preserve functional groups. Because they do not distinguish whether a chemical bond belongs to a functional group, they ruin the molecular structures that are vital for determining molecule properties. (2) Two chemical-reaction-aware methods behave differently. As RECAP has more conservative rules, it produces a vocabulary at least one magnitude larger than its counterparts.

Because BRICS shows a better balance among these principles, we adapt it to serve our pre-trained model. Furthermore, because over 60% of the out-of-vocabulary problem, an essential factor affecting the pre-trained model quality [6], is caused by the combinations between rings and other structures through single bonds, we add an extra rule to BRICS: cut all single bonds attached to a ring.

**Details of Discretization** To stabilize the training process, we discretize two continuous variables, i.e., translation vector and rotation quaternion. Specifically, we map them into two discrete spaces.

For the translation vector, its discrete space is grids in 3D space (see Figure 1a). Briefly speaking, we represent this continuous vector with the discrete index of the grid. We represent the $i$-th translation bin as the coordinate of its centre $t_i^{\text{bin}} \in \mathbb{R}^3$. The discretization of any continuous translation operator $t \in \mathbb{R}^3$ can be computed by $\arg\min_i \|t_i^{\text{bin}}, t\|_2$. The total number of grids is $21,952$.

Because the rotation axis and rotation angle can determine the rotation quaternion, the discrete space of the quaternion contains two parts. As shown in Figure 1b, we first enumerate rotation axes $(x, y, z)$ in 3D space, then for each axis, we list the rotation angle every $\theta$ degrees. We represent the $i$-th rotation bin (stands for rotating $\theta_i$ degrees around an axis $(x_i, y_i, z_i)$) as a quaternion

**Algorithm 1** Pocket-based Sketching
___
**Input:** a function $\lambda$ measuring the volume of a shape, a threshold $t$, a step size $\alpha$
a pocket shape $S_p$ located at $C_p$, a seed shape $S_s$ located at $C_s$ where $\lambda\left(\bigcap(S_p, S_s)\right) = 0$
**Output:** The sampled molecule shape $S_d$.
    **while** True **do**
        $V \leftarrow \lambda\left(\bigcap(S_p, S_s)\right)$
        **if** $V \geq t$ **then**
            $S_d \leftarrow \bigcap(S_p, S_s)$
            **break**
        **end if**
        $C_s = \alpha C_p + (1 - \alpha)C_s$
    **end while**
___

Table 1: Comparison of different tokenizing methods. Coverage is the percentage of novel molecules that the vocabulary can handle.

| Method | Coverage | Vocabulary Size | Case |
| --- | --- | --- | --- |
| Atom-based | 100.0 | 49 | |
| Cut Single Bond | 74.3 | 16,920 | |
| RECAP | 40.1 | 289,188 | |
| BRICS | 61.2 | 49,339 | |
| Ours | 70.4 | 23,896 | |

$q_i^{\text{bin}} = (x_i, y_i, z_i, \theta_i) \in \mathbb{R}^4$. The discretization of any continuous rotation operator $q \in \mathbb{R}^4$ can be computed by $\arg\min_i \|q_i^{\text{bin}}, q\|_2$. The total number of bins is $8,763$. To be precise, we enumerate 363 axes and 24 angles (i.e., 15 degrees per angle).

Another reason for the discretization is to avoid the discontinuity of quaternions when optimizing them [7, 8]. Specifically, there are several approaches to parameterize the rotation operator: quaternion [9], euler-angle [10], and SO(3) group (i.e., the rotation matrix) [11]. Quaternion and euler-angle are sometimes ambiguous and discontinuous. For example, the rotation operator is periodic, rotating $180°$ is equal to rotating $-180°$ and rotating $179.9°$ is very close to rotating $-179.9°$. Without discretization, we will excessively penalize the model according to the mean-square-error (i.e., $[179.9 - (-179.9)]^2 = 359.8^2$). While in this paper, we convert a regression problem into a classification one with discretization, thus avoiding the above issues. Instead of discretizing quaternions, AlphaFold [12] avoids such issues by regarding the quaternion as an intermediate variable and not optimizing the quaternion directly. We leave adopting AlphaFold's strategy for future work.

**Greedy Algorithm for Connecting Fragments** Algorithm 2 is responsible for converting the separated fragments into a complete 3D molecule. To be specific, we first place all the fragments in 3D space according to the predictions of SHAPE2MOL. Then, for each time, we greedily chose the

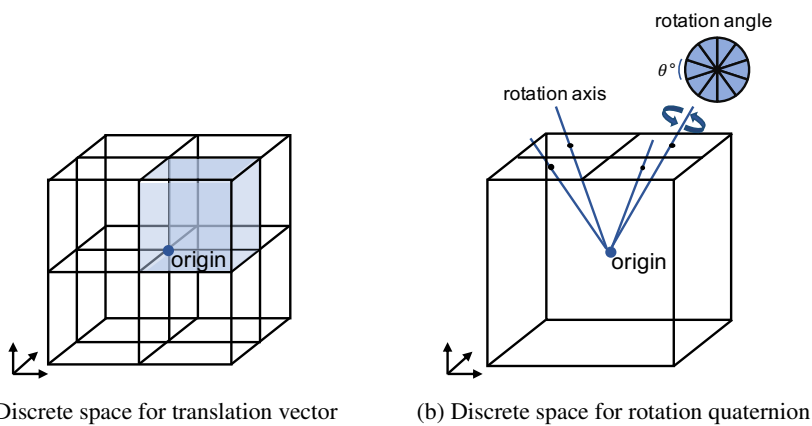

(a) Discrete space for translation vector  (b) Discrete space for rotation quaternion

Figure 1: Discretization of translation vector and rotation quaternion. Both origins locate at the centroid of the fragment.

two closest breakpoints from different fragments and connect the fragments through the breakpoints. By repeating the process, the fragments get larger and larger. When there are not enough breakpoints to connect, we return the largest fragment as the final molecule. For potential residual breakpoints, we attach carbon atoms to them for molecular validity.

---

**Algorithm 2** Greedy Algorithm for Connecting Fragments

---

**Input:** Fragment sequence $S = [(C_i, P_i, R_i)]_{i=0}^{l_s}$, fragment vocabulary $V$.
**Output:** A whole molecule $M$ formed by the fragments.
  $B \leftarrow \{\}$
  $F \leftarrow \{\}$
  **for** $i \leftarrow 0$ to $l_s$ **do**
    $(C_i, P_i, R_i) \leftarrow S[i]$
    $f \leftarrow$ query_vocabulary$(V, C_i)$
    $f \leftarrow$ rotate(translate($f, P_i$), $R_i$)     ▷ Place the fragment in 3D space
    $b \leftarrow$ get_breakpoints$(f)$     ▷ Return all the breakpoints in current fragment
    $F \leftarrow F \cup \{f\}$
    $B \leftarrow B \cup \{b\}$
  **end for**
  **while** $|B| \geq 2$ **do**
    $b_1, b_2 \leftarrow$ get_breakpoints$(B)$     ▷ Chose two nearest breakpoints from different fragments
    **if** $b_1$ is None or $b_2$ is None **then**
      **break**
    **end if**
    $f_1 \leftarrow$ get_fragment$(F, b_1)$     ▷ Retrieve the corresponding fragment of the breakpoint
    $f_2 \leftarrow$ get_fragment$(F, b_2)$
    $M_{partial} \leftarrow$ attach_fragments$(f_1, b_1, f_2, b_2)$     ▷ Connect fragments through the breakpoints
    $B \leftarrow B \setminus \{b_1, b_2\}$
    $F \leftarrow F \setminus \{f_1, f_2\}$
    $F \leftarrow F \cup \{M_{partial}\}$
  **end while**
  $M \leftarrow$ get_largest_fragment$(F)$     ▷ Chose the largest fragment as the output molecule
  **if** $|B| \neq 0$ **then**
    $M \leftarrow$ attach_carbon$(M, B)$     ▷ Handle the remaining breakpoints by attaching carbon atoms
  **end if**

---

### 1.3 More Details of SHAPE2MOL

**Output of Shape Encoder**   The output of the shape encoder is the continuous representation of each 3D patch, which contain the geometric information of input molecular shape. It will serve as the context of the decoder to constrain the shape of generated molecules.

**Input of Shape Decoder**   The input of the shape decoder at the time step $t$ is the fragment category, rotation quaternion, and translation vector from the output of the previous step $t - 1$. We use them to tell the model how exactly a fragment is placed in 3D space so that the model can generate the next fragment connected with it. The output of the shape encoder is fed into the decoder as the geometric context, through the cross-attention module.

**Hyperparameters of SHAPE2MOL**   Both encoder and decoder of SHAPE2MOL have 12 stacked 8-head Transformer layers, whose model size is 1024 and feedforward dimension is 4096. We use ReLu as the activation function and set the position embedding learnable in the training process.

For the encoder, following liGAN, we set the resolution of the voxelized shape to 0.5Å and the side length of the spanned cube to 14Å. The size of 3D patches is $4 \times 4 \times 4$, which allows us to handle a large number of voxels. Specifically, we reshape $21,952$ voxels into 343 3D patches in the paper.

The decoder has 3 different embeddings, denoting the fragment category, the discretized rotation, and the discretized translation, respectively. Similarly, we also use 3 different output heads to predict the category, rotation, and translation. The output heads share their weights with the embeddings. The total number of model parameters is 650M.

## 2 Experiments

### 2.1 Metrics

- Uniqueness (**Uniq**) is the percentage of unique molecules among all generated results.
- Novelty (**Nov**) is the percentage of generated molecules with Tanimoto similarity [13] less than 0.4 compared to its nearest neighbor in existing ligands [14].
- Diversity (**Div**) is the internal diversity of the generated molecules and calculated as $\frac{2}{n(n-1)} \sum_{x \neq x'} 1 - sim(x, x^{'})$.
- Success rate (**Succ**) is the percentage of generated molecules that pass the predefined thresholds for the desired properties. Specifically, we based on a widely used rule of thumb [15] to set the thresholds as $\text{QED} \geq 0.25$, $\text{SA}_{\text{score}} \geq 0.59$ and $\text{Vina}_{\text{score}} \leq -8.18$ kcal/mol, where QED is the index of drug-likeness [16] and $\text{SA}_{\text{score}}$ is the index of synthetic accessibility [17].
- Product (**Prod**) is the product of the four metrics above and serves as a comprehensive evaluation.
- Median Vina Score (**Median**) is the median of $\text{Vina}_{\text{score}}$. $\text{Vina}_{\text{score}}$ is the binding energy evaluation using the global energy optimization algorithm in Vina [18], which can reflect the binding affinity between proteins and molecules.

### 2.2 More Ablation Study of Generating

**Model Variants**   As shown in Figure 3, we study two variants in the model design: discretization and robust training. The discretization consistently improves the result, while the continuous variant is worse. A possible reason is a non-linear relationship between quaternions and rotation angles. [1] We observe minor improvement for finer granularity and leave it for future exploration. We also study the ability of handling shape noise caused by a possibly large pocket size. The model trained with shape noise (i.e. robust training) shows better performance when test data are noisy, while the performance under standard training drops clearly.

---

[1]For example, if a molecule rotates $90°$ along an axis, the mean squared error of quaternion is 1; when the angle is $10°$, the error is 0.98.

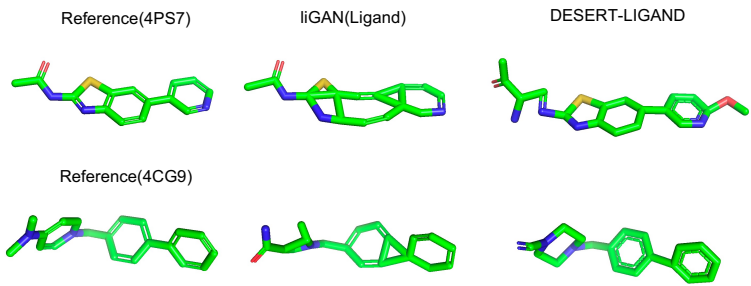

Figure 2: More cases from liGAN and DESERT.

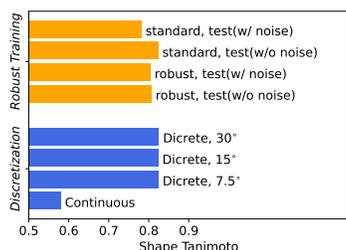

Figure 3: Comparison of discretization and robust training.

Table 2: Comparison of difference decoding strategy.

| Sampling Method | Div | Prod | Median |
|---|---|---|---|
| Beam Search (Beam=10) | 0.70 | 0.00 | -5.87 |
| Greedy Decoding | 0.65 | 0.00 | -5.83 |
| Top K (K=10) | 0.91 | 0.04 | -5.99 |
| Top P (P=0.95) | 0.93 | 0.06 | -6.01 |
| + post-processing | 0.92 | 0.17 | -7.34 |

Table 3: Combine protein chemical information.

| Method | Prod | Median |
|---|---|---|
| DESERT-POCKET | 0.55 | -9.32 |
| + chemical(Weak) | 0.53 | -9.19 |
| + chemical(Strong) | 0.52 | -9.03 |

Table 4: Comparison of difference seed shape.

| Seed Shape | Succ | Prod | Median |
|---|---|---|---|
| None | 82.0 | 0.74 | -8.85 |
| Sphere | 75.5 | 0.68 | -9.13 |
| Molecules | 61.0 | 0.55 | -9.32 |

**Decoding Strategy** We replace the Nucleus sampling method, i.e., Top P sampling, in the decoding strategy with Beam Search [19], Greedy Decoding [20], and Top K sampling [20] to study the influence of different sampling methods on the performance of DESERT. As shown in Table 2, Beam Search and Greedy Decoding are poor at providing diverse molecules, which suggests that the output probability is concentrated. They can only get several fragments with high output probability. When low probability fragments are obtained, the ranking mechanism in Beam Search drops them in the subsequent decoding step. Due to their internal annealing mechanism, both Top K and Top P can provide relatively diverse results.

**Chemical Information Driven Design** We also explore the potential of integrating chemical information of proteins into drug design. Briefly speaking, based on hydrogen bond acceptor-donor rules [21], we put the fragment with more hydrogen atoms into the pocket region with more oxygen, nitrogen, and fluorine atoms. The results in Table 3 show that the strategy does not achieve better performance when we increase the effect of chemical information on the output probability. The reason might be not considering detailed chemical information, like the bond length. We leave a more thoughtful method for combining the chemical information for future work.

**Atom-based Pre-training** We pre-train an atom-based SHAPE2MOL to analyze the contribution of our fragment-based generation style. Specifically, we keep most pre-training processes the same but replace our tokenizing method with the atom-based strategy (both mentioned in Section 1.2). As drug molecules mainly consist of carbon atoms, the atom-based SHAPE2MOL ignores input shapes and keeps outputting carbon atoms. The atom-based variant can not design any valid molecules, which indicates the fragment-based style is crucial for our pre-training process.

5

**Invariance of SHAPE2MOL** Following liGAN, when training SHAPE2MOL, we randomly rotate and translate the molecules to make our model have rotation and translation invariance ability. We compare the similarity of generated molecules based on different or the same molecular shapes (both are randomly rotated and translated). The similarity rises from 0.092 to 0.508, which shows that with the same molecular shape as input, the model produces similar molecules.

### 2.3 More Ablation Study of Sketching

**Seed Shape Type** We evaluate the influence of different seed shape in Table 4. We find that directly using the pocket as the shape may contribute to suboptimal result in providing highly active molecules. Through the sketching stage, we obtain more realistic and diverse molecular shapes. These shapes benefit the molecular activity but result in more complex structures and lower synthetic accessibility in generated molecules.

## 3 Discussion of Related Work

### 3.1 Shape-based Drug Design

Here we discuss the relationship between our method and existing shape-based drug design [22] approaches. Some previous work designed new drugs based on the shape of a known ligand. Traditional approaches work in a retrieval way – finding molecules whose shape is most similar to a known one [23, 24]. Modern deep learning models can decode a molecule from its shape [1, 25]. Such ligand-based generation can not generalize to unseen pockets. Luo et al. [2] directly generates molecules from the pocket shape, which is mostly close to our work. However, our model works in a fragment-based fashion, while theirs works in an atomic way. What makes our model especially different is that we utilize the power of the pre-training model to make pocket-based drug design more promising.

### 3.2 Tokenization and Linearization

Here we discuss how the tokenization and linearization procedures relate to fragment-based drug design (FBDD) [26, 27].

**Fragment-Based Drug Design** Briefly, there are two approaches in FBDD: growing the fragment synthetically to a proximal binding site or by linking two fragments together [27]. Our method can be classified as the previous type since the linearization generates a molecule in a one-by-one fashion.

**Tokenization** The procedure is carefully designed for deep generative models to avoid loops and preserve functionalities, which is in spirit to the principle of [3]. Some FBDD work, such as [28] works in a discriminative approach. Thus there are not many constraints when cutting the molecules. Podda et al. [29] uses the SMILES-fragment rather than a real molecule-fragment. Thus it can not utilize rich structured features.

**Linearization** The procedure aims at traversing (or generating) a structured object in a left-to-right approach [30], which is tractable and scalable. It is borrowed from the area of computational linguistics, more specifically, syntactic parsing and structured generation [31, 32]. For micro-molecules, linearization (such as SMILES [33]) has been adopted for several decades. There is a line of research work for SMILES-based generation [34, 35, 36]. Similarly, in the area of macro-molecule, Huang et al. [37] designs a linear structure for estimating the likelihood of the structure of RNA.

**Comparison** Compared with traditional linearized sequences such as SMILES [33], our method utilizes structural information to segment the molecules to preserve their functionality. Compared with topological generation based on a graph [3, 38, 39], our method is more scalable to big data since generating the variable graph topology is not friendly to large-batch training in neural networks.

## 4 More Cases

We show more cases from liGAN and DESERT in Figure 2. Similarly, liGAN struggles to produce realistic molecular structures, while DESERT generates better results.

## References

[1] Tomohide Masuda, Matthew Ragoza, and David Ryan Koes. "Generating 3D Molecular Structures Conditional on a Receptor Binding Site with Deep Generative Models". In: *CoRR* abs/2010.14442 (2020). arXiv: 2010.14442. URL: https://arxiv.org/abs/2010.14442.

[2] Shitong Luo et al. "A 3D Generative Model for Structure-Based Drug Design". In: *Advances in Neural Information Processing Systems* 34 (2021).

[3] Wengong Jin, Regina Barzilay, and Tommi Jaakkola. "Junction tree variational autoencoder for molecular graph generation". In: *International conference on machine learning*. PMLR. 2018, pp. 2323–2332.

[4] Xiao Qing Lewell et al. "Recap retrosynthetic combinatorial analysis procedure: a powerful new technique for identifying privileged molecular fragments with useful applications in combinatorial chemistry". In: *Journal of chemical information and computer sciences* 38.3 (1998), pp. 511–522.

[5] Jrg Degen et al. "On the Art of Compiling and Using'Drug-Like'Chemical Fragment Spaces". In: *ChemMedChem: Chemistry Enabling Drug Discovery* 3.10 (2008), pp. 1503–1507.

[6] Wen Tai et al. "exBERT: Extending pre-trained models with domain-specific vocabulary under constrained training resources". In: *Findings of the Association for Computational Linguistics: EMNLP 2020*. 2020, pp. 1433–1439.

[7] Yi Zhou et al. "On the continuity of rotation representations in neural networks". In: *Proceedings of the IEEE/CVF Conference on Computer Vision and Pattern Recognition*. 2019, pp. 5745–5753.

[8] Luca Falorsi et al. "Explorations in homeomorphic variational auto-encoding". In: *arXiv preprint arXiv:1807.04689* (2018).

[9] Wikipedia contributors. *Quaternion — Wikipedia, The Free Encyclopedia*. https://en.wikipedia.org/w/index.php?title=Quaternion&oldid=1112663286. [Online; accessed 4-October-2022]. 2022.

[10] Wikipedia contributors. *Euler angles — Wikipedia, The Free Encyclopedia*. https://en.wikipedia.org/w/index.php?title=Euler_angles&oldid=1112943781. [Online; accessed 4-October-2022]. 2022.

[11] Wikipedia contributors. *3D rotation group — Wikipedia, The Free Encyclopedia*. https://en.wikipedia.org/w/index.php?title=3D_rotation_group&oldid=1101034704. [Online; accessed 4-October-2022]. 2022.

[12] John Jumper et al. "Highly accurate protein structure prediction with AlphaFold". In: *Nature* 596.7873 (2021), pp. 583–589.

[13] David Rogers and Mathew Hahn. "Extended-connectivity fingerprints". In: *Journal of chemical information and modeling* 50.5 (2010), pp. 742–754.

[14] Marcus Olivecrona et al. "Molecular de-novo design through deep reinforcement learning". In: *Journal of cheminformatics* 9.1 (2017), pp. 1–14.

[15] Oleg Ursu et al. "DrugCentral 2018: an update". In: *Nucleic acids research* 47.D1 (2019), pp. D963–D970.

[16] G Richard Bickerton et al. "Quantifying the chemical beauty of drugs". In: *Nature chemistry* 4.2 (2012), pp. 90–98.

[17] Peter Ertl and Ansgar Schuffenhauer. "Estimation of synthetic accessibility score of drug-like molecules based on molecular complexity and fragment contributions". In: *Journal of cheminformatics* 1.1 (2009), pp. 1–11.

[18] Oleg Trott and Arthur J Olson. "AutoDock Vina: improving the speed and accuracy of docking with a new scoring function, efficient optimization, and multithreading". In: *Journal of computational chemistry* 31.2 (2010), pp. 455–461.

[19] Markus Freitag and Yaser Al-Onaizan. "Beam search strategies for neural machine translation". In: *arXiv preprint arXiv:1702.01806* (2017).

[20] Ari Holtzman et al. "The curious case of neural text degeneration". In: *arXiv preprint arXiv:1904.09751* (2019).

[21] Roderick E Hubbard and Muhammad Kamran Haider. "Hydrogen bonds in proteins: role and strength". In: *eLS* (2010).

[22] Rob LM Van Montfort and Paul Workman. "Structure-based drug design: aiming for a perfect fit". In: *Essays in biochemistry* 61.5 (2017), pp. 431–437.

[23] Ashutosh Kumar and Kam YJ Zhang. "Advances in the development of shape similarity methods and their application in drug discovery". In: *Frontiers in chemistry* 6 (2018), p. 315.

[24] Camila Cardoso Santos et al. "Drug screening using shape-based virtual screening and in vitro experimental models of cutaneous Leishmaniasis". In: *Parasitology* 148.1 (2021), pp. 98–104.

[25] Miha Skalic et al. "Shape-based generative modeling for de novo drug design". In: *Journal of chemical information and modeling* 59.3 (2019), pp. 1205–1214.

[26] Philine Kirsch et al. "Concepts and core principles of fragment-based drug design". In: *Molecules* 24.23 (2019), p. 4309.

[27] Christopher W Murray and David C Rees. "The rise of fragment-based drug discovery". In: *Nature chemistry* 1.3 (2009), pp. 187–192.

[28] Harrison Green, David R Koes, and Jacob D Durrant. "DeepFrag: a deep convolutional neural network for fragment-based lead optimization". In: *Chemical Science* 12.23 (2021), pp. 8036–8047.

[29] Marco Podda, Davide Bacciu, and Alessio Micheli. "A deep generative model for fragment-based molecule generation". In: *International Conference on Artificial Intelligence and Statistics*. PMLR. 2020, pp. 2240–2250.

[30] Yijia Liu et al. "Transition-based syntactic linearization". In: *Proceedings of the 2015 Conference of the North American Chapter of the Association for Computational Linguistics: Human Language Technologies*. 2015, pp. 113–122.

[31] Yoon Kim et al. "Unsupervised recurrent neural network grammars". In: *arXiv preprint arXiv:1904.03746* (2019).

[32] Oriol Vinyals et al. "Grammar as a foreign language". In: *Advances in neural information processing systems* 28 (2015).

[33] David Weininger. "SMILES, a chemical language and information system. 1. Introduction to methodology and encoding rules". In: *Journal of chemical information and computer sciences* 28.1 (1988), pp. 31–36.

[34] Matt J Kusner, Brooks Paige, and José Miguel Hernández-Lobato. "Grammar variational autoencoder". In: *International conference on machine learning*. PMLR. 2017, pp. 1945–1954.

[35] Hanjun Dai et al. "Syntax-directed variational autoencoder for structured data". In: *arXiv preprint arXiv:1802.08786* (2018).

[36] Seokho Kang and Kyunghyun Cho. "Conditional molecular design with deep generative models". In: *Journal of chemical information and modeling* 59.1 (2018), pp. 43–52.

[37] Liang Huang et al. "LinearFold: linear-time approximate RNA folding by 5'-to-3'dynamic programming and beam search". In: *Bioinformatics* 35.14 (2019), pp. i295–i304.

[38] Binghong Chen et al. "Molecule Optimization by Explainable Evolution". In: *International Conference on Learning Representations*. 2020.

[39] Yutong Xie et al. "Mars: Markov molecular sampling for multi-objective drug discovery". In: *arXiv preprint arXiv:2103.10432* (2021).



\end{CJK*}
\end{document}